\documentclass[seceq]{ptptex}
\usepackage{wrapft}
\usepackage{graphicx}
\newcommand{\ket}[1]{| {#1} \rangle}     %%
\newcommand{\rbra}[1]{( {#1} |}     %%
\newcommand{\rket}[1]{| {#1} )}     %%
\newcommand{\rdket}[1]{|\!| {#1} )}     %%
\newcommand{\maru}[1]{\stackrel{{\tiny\circ}} {#1}}
\newcommand{\wtilde}[1]{\widetilde{#1}} %%

\newcommand{\ovl}[1]{\overline{#1}}

\def\beq{\begin{eqnarray}}
\def\eeq{\end{eqnarray}}
\def\bsub{\begin{subequations}}
\def\esub{\end{subequations}}
\def\b{\begin{equation}}
\title{%        %You can use \\ for explicit line-break
The BCS-Bogoliubov and the $su(2)$-Algebraic Approach to the Pairing Model in Many-Fermion System
}
\subtitle{
The Quasiparticle in the Conservation of the Fermion Number
}    %use this when you want a subtitle

\author{%       %Use \sc for the family name
Yasuhiko {\sc Tsue},$^{1}$ 
Constan\c{c}a {\sc Provid\^encia},$^{2}$ 
Jo\~ao da {\sc Provid\^encia}$^{2}$ and 
Masatoshi {\sc Yamamura}$^{3}$  
}

\inst{%         %Affiliation, neglected when [addenda] or [errata]
$^{1}$Physics Division, Faculty of Science, Kochi University, Kochi 780-8520, 
Japan\\
$^{2}$Departamento de F\'{i}sica, Universidade de Coimbra, 3004-516 Coimbra, 
Portugal\\
$^{3}$Department of Pure and Applied Physics, 
Faculty of Engineering Science,\\
Kansai University, Suita 564-8680, Japan
\\

}

\recdate{%      %Editorial Office will fill in this.
\today
}

\abst{%       %this abstract is neglected when [addenda] or [errata]
The relation between two approaches to the $su(2)$-algebraic many-fermion 
model is discussed: (1) the BCS-Bogoliubov approach in terms of the 
use of the quasiparticles representing all the degrees of freedom except 
those related to the Cooper-pairs and (2) the conventional 
algebraic approach in terms of the use of the minimum weight states, from which 
the Cooper-pairs are excluded. 
In order to arrive at the goal, the idea of the quasiparticles is 
brought up in the conservation of the fermion number. 
Under the $c$-number replacement for the three $su(2)$-generators, the 
quasiparticles suggested in this paper are reduced to those in the 
BCS-Bogoliubov approach. 
It is also shown that the two approaches are equivalent 
through the $c$-number replacement. 
Further, a certain modification of the BCS-Bogoliubov approach is discussed. 
}

\begin{document}

\maketitle

\section{Introduction}

Under the name of the pairing model, the $su(2)$-algebraic many-fermion model gives us a schematic 
understanding for the pairing correlation, one of the most fundamental correlations in nuclei and, recently, 
the present authors have published two papers on this model.\cite{1} 
The present paper is a continuation of these two papers, in which we mentioned the reason 
why the pairing model must be investigated in spite of being a rather old-fashioned problem. 
Therefore, in the present paper, we will not repeat the general discussion on the reason. 
Hereafter, these two papers will be referred to as (I) collectively.

As is well known, the study of the pairing correlation in nuclei has occupied a central part 
in theoretical nuclear physics. 
In particular, immediately after the BCS-Bogoliubov theory\cite{2} for the 
superconductivity appeared, many-body theoretical study of the pairing correlation in nuclei has 
remarkably developed. 
We will mention a rough sketch of the study at this period. 
In 1958, Bohr, Mottelson and Pines suggested that the medium and heavy nuclei are in a superconductivity phase.\cite{3}
Following this suggestion, in 1959, Belyaev applied the BCS-Bogoliubov theory to the description 
of low-lying collective vibrational states of spherical nuclei.\cite{4}
His idea is based on the adiabatic approximation combined with the cranking model 
by Inglis.\cite{5} 
In this work, the ground states are regarded as the condensed Cooper-pair states in spherical shape and 
the excited states are treated in terms of appropriate superposition of the quasiparticles 
induced by correlations which do not belong to the pairing correlation. 
As a possible extension of the method of the adiabatic approximation, 
Marumori, Arvieu \& Veneroni and Baranger, independently, proposed the so-called quasiparticle random phase 
approximation in 1960.\cite{6} 
With the aid of this method, we can understand the structure of the correlated ground states and 
non-collective excited states orthogonal to the collective excited states. 
On the other hand, in 1961, an extension of the BCS-Bogoliubov theory, 
which can be applied to the deformed shape, was discussed by Baranger.\cite{7} 
It is called the Hartree-Bogoliubov method. 
On the basis of this method, 1962, the idea of Belyaev was applied to the case of 
deformed nuclei by Marumori, one of the present authors (M. Y.) and Bando.\cite{8} 
The above is a rough sketch on the study of the BCS-Bogoliubov approach to nuclear physics at the early stage. 
The ideas proposed at this stage have been spread to the studies of other various fields in 
nuclear physics. 
As was mentioned above, the BCS-Bogoliubov theory occupies an important position and for any theory 
related to the pairing model, the relevance to the BCS-Bogoliubov theory should be mentioned. 
But, in (I), we did not contact with this theory.

It is also well known that the pairing correlation is successfully described by the $su(2)$-algebra 
consisting of three generators. 
They are the Cooper-pair creation and annihilation operators and the hermitian operator related to 
the total fermion number. 
In the conventional algebraic approach, the orthogonal basis for the irreducible representation 
is composed of the states obtained by successive operation of the 
Cooper-pair creation operator on the minimum weight states. 
They are expressed in terms of all the degrees of freedom except the Cooper-pairs and the 
total number of the fermions contained in these states is called the 
seniority number.
Symbolically, the states for the orthogonal basis in the conventional algebraic approach 
can be expressed in the form ${\wtilde C}_S^*{\wtilde D}_{A}^*\rket{0}$. 
Here, ${\wtilde C}_S^*$, ${\wtilde D}_A^*$ and $\rket{0}$ denote the operator composed of the 
successive operation of the Cooper-pair creation, the operator forming the minimum weight state 
and the vacuum of the original fermions, respectively. 
Explicit forms will be shown in Eqs.(\ref{2-24}) and (\ref{2-21}), respectively. 
Therefore, it may be the starting point for the $su(2)$-algebraic many-fermion model to 
determine the minimum weight state. 
For this task, it is unavoidable to express all the degrees of freedom except the 
Cooper-pair in terms of functions of the original fermion operators. 
In (I), the present authors have presented a possible systematic method how to determine 
the minimum weight state. 
It is formulated in the framework of the conservation of the fermion number. 
On the other hand, the orthogonal basis in the BCS-Bogoliubov approach is constructed in the 
form of the quasiparticles created on the condensed Cooper-pair state 
and, then, the seniority number corresponds to the number of these quasiparticles. 
The states for the orthogonal basis in the BCS-Bogoliubov approach can be expressed, symbolically, 
in the form ${\wtilde D}_B^*{\wtilde C}_D^*\rket{0}$. 
Here, ${\wtilde C}_D^*$ and ${\wtilde D}_B^*$ denote the operator for the condensed Cooper-pair state and 
the operator as function of the quasiparticles created on the state ${\wtilde C}_D^*\rket{0}$, 
respectively, which will be shown also in Eqs.(\ref{2-4}) and (\ref{2-8}), respectively. 
The condensed Cooper-pair state is in the framework of the non-conservation of the fermion number. 
As may be clear from the above argument, the contrast between the BCS-Bogoliubov and 
the $su(2)$-algebraic approach is remarkable: 
(1) the frameworks are non-conservation and conservation of the fermion number, 
(2) all the degrees of freedom except the Cooper-pairs are expressed in terms of the 
quasiparticles and the minimum weight state and (3) the order of the operation of the Cooper-pair 
creation operator and the operators related to all the degrees of freedom except 
the Cooper-pairs is reverse, i.e., 
${\wtilde D}_B^*{\wtilde C}_D^*\rket{0}$ and ${\wtilde C}_S^*{\wtilde D}_A^*\rket{0}$. 
Therefore, it may be interesting and important to investigate if both approaches are 
equivalent or not. 
Hereafter, the $su(2)$-algebraic and the BCS-Bogoliubov approach will be abbreviated as 
A and B, respectively and also the approach presented in this paper as C.

The main aim of this paper is to show that the approach A is reduced to B under the $c$-number replacement of the 
$su(2)$-generators. 
For completing this task, we will propose the approach C. 
In B, regarding the condensed Cooper-pair state as the vacuum, the quasiparticle is introduced. 
In C, the quasiparticle is defined in the framework of the conservation of the fermion number. 
In the state which corresponds to the condensed Cooper-pair state, the number of the 
Copper-pairs is fixed. 
With the use of the new quasiparticle, the minimum weight state is constructed. 
It is equivalent to the minimum weight state expressed in terms of the original fermion 
operators, which is presented in (I). 
As was already mentioned, the states composing the orthogonal basis in C (A) are of the 
form constructed by operating the Cooper-pair operators on the minimum weight state. 
Symbolically, we have the relation ${\wtilde C}_S^*{\wtilde D}_C^*\rket{0}={\wtilde C}_S^*{\wtilde D}_A^*\rket{0}$. 
Here, ${\wtilde D}_C^*$ denotes the operator for the minimum weight state in (C), the explicit form 
of which will be given in Eq.(\ref{4-11}). 
Therefore, in order to discuss the connection to B, the orthogonal basis must be rewritten to the form in which 
operation of the Cooper-pair operator and the quasiparticle is reversed. 
This statement can be expressed in terms of ${\wtilde C}_S^*{\wtilde D}_C^*\rket{0}={\wtilde D}_C'^*{\wtilde C}_S^*\rket{0}$. 
Here, ${\wtilde D}_C'^*$ denotes the operator obtained after the reversion. 
The explicit form of ${\wtilde D}_C'^*$ will be given in Eq.(\ref{5-8}). 
For this task, we present a certain idea. 
After rewriting, we replace the $su(2)$-generators with the $c$-numbers. 
Then, the states composing the orthogonal basis is reduced to those given in B. 
Thus, we have the following conclusion: 
The approach A is equivalent to B under the condition that the $su(2)$-generators are replaced with the 
$c$-numbers. 
Of course, the conclusion is obtained by the media of the approach C developed in this paper. 
However, we have a certain problem on the $c$-number replacement. 
We also investigate the relation between the approach C and a revised version of the 
approach B presented by Suzuki and Matsuyanagi\cite{9} and, later, by Kuriyama and one of the 
present authors (M. Y.).\cite{10}

In next section, the $su(2)$-algebraic model and its two approaches are 
recapitulated. 
In particular, the main part of (I) is summarized. 
In \S 3, as the approach C, the definition of the quasiparticle is given 
in the framework of conservation of the fermion number. 
It is shown that, under the $c$-number replacement, the quasiparticle defined in \S 3 reduces to 
that in the approach B. 
In \S 4, the minimum weight state in terms of the quasiparticles is presented and it is shown that the states 
constructed on the minimum weight state are equivalent to those given in (I). 
Section 5 is devoted to rewriting the set of the states given 
in \S 4. 
Further, it is shown that, under the $c$-number replacement for the $su(2)$-generators, the set of the states is 
reduced to the orthogonal basis in the approach B. 
Finally in \S 6, mainly, a certain problem appearing in the $c$-number replacement 
is discussed.

\section{The $su(2)$-algebraic many-fermion model and two approaches}

For the convenience of later discussion, we will recapitulate the model 
described in this paper briefly including the interpretation of the 
notations. 
We treat a many-fermion system in which the constituents are under the 
pairing correlation and confined in $4\Omega_0$ single-particle states. 
Here, $\Omega_0$ denotes an integer or half-integer. 
Since $4\Omega_0$ is an even-number, all single-particle states are divided into equal 
parts $P$ and ${\ovl P}$.
Therefore, as a partner, each single-particle state belonging to $P$ can find 
a single-particle state in ${\ovl P}$. 
We express the partner of the state $\alpha$ belonging to $P$ as ${\ovl \alpha}$ and 
fermion operators in $\alpha$ and ${\ovl \alpha}$ are denoted as 
$({\tilde c}_{\alpha}, {\tilde c}_{\alpha}^*)$ and 
$({\tilde c}_{{\bar \alpha}}, {\tilde c}_{{\bar \alpha}}^*)$, respectively. 
As the generators ${\wtilde S}_{\pm,0}$, we adopt the following form: 
\beq\label{2-1}
{\wtilde S}_+=\sum_{\alpha}s_{\alpha}{\tilde c}_{\alpha}^*{\tilde c}_{\bar \alpha}^*\ , \quad
{\wtilde S}_-=\sum_{\alpha}s_{\alpha}{\tilde c}_{\bar \alpha}{\tilde c}_{\alpha}\ , \quad
{\wtilde S}_0=\frac{1}{2}\sum_{\alpha}({\tilde c}_{\alpha}^*{\tilde c}_{\alpha}+
{\tilde c}_{\bar \alpha}^*{\tilde c}_{\bar \alpha})-\Omega_0\ .
\eeq
The symbol $s_{\alpha}$ denotes a real number satisfying $s_{\alpha}^2=1$. 
The sum $\sum_{\alpha}$ ($\sum_{\bar \alpha}$) is 
carried out in all single-particle states in $P$ (${\ovl P}$). and 
we have $\sum_{\alpha}1=2\Omega_0$ ($\sum_{\bar \alpha}1=2\Omega_0$). 
Conventionally, in the case of the pairing correlation, ${\wtilde S}_+$ is 
called the Cooper-pair creation operator. 
The commutation relation for ${\wtilde S}_{\pm,0}$ is given as 
\beq\label{2-2}
[\ {\wtilde S}_+\ , \ {\wtilde S}_-\ ]=2{\wtilde S}_0\ , \qquad
[\ {\wtilde S}_0\ , \ {\wtilde S}_{\pm}\ ]=\pm{\wtilde S}_{\pm}\ .
\eeq
The Casimir operator ${\wtilde {\mib S}}^2$, which commutes with 
${\wtilde S}_{\pm,0}$, is defined as 
\beq\label{2-3}
{\wtilde {\mib S}}^2={\wtilde S}_+{\wtilde S}_-+{\wtilde S}_0^2-{\wtilde S}_0\ .
\eeq
The above is the recapitulation of the present model.

Let us discuss the BCS-Bogoliubov approach, which is called the approach B in this 
paper. 
It starts in the following state: 
\beq\label{2-4}
\rket{\phi}=\exp\left(\frac{v}{u}{\wtilde S}_+\right)\rket{0}\ . \quad
\left({\wtilde C}_{D}^*=\exp\left(\frac{v}{u}{\wtilde S}_+\right)\right)
\eeq
Here and hereafter, we omit the normalization constant for any state. 
The parameters $u$ and $v$ are positive and complex, respectively and 
obey the condition 
\beq\label{2-5}
u^2+|v|^2=1\ .
\eeq
We call $\rket{\phi}$ as the condensed Cooper-pair state. 
Clearly, $\rket{\phi}$ is not eigenstate of fermion number operator, 
${\wtilde N}=\sum_{\alpha}({\tilde c}_{\alpha}^*{\tilde c}_{\alpha}
+{\tilde c}_{\bar \alpha}^*{\tilde c}_{\bar \alpha})$ and it is the vacuum 
of the following operators $({\maru d}_{\alpha}, {\maru d}{}_{\alpha}^*)$ and 
$({\maru d}_{\bar \alpha}, {\maru d}{}_{\bar \alpha}^*)$: 
\bsub\label{2-6}
\beq
& &{\maru d}_{\alpha}\rket{\phi}={\maru d}_{\bar \alpha}\rket{\phi}=0\ , 
\label{2-6a}\\
& &{\maru d}{}_{\alpha}^*=u{\tilde c}_{\alpha}^*-v^*s_{\alpha}{\tilde c}_{\bar \alpha}\ , 
\qquad\qquad
{\tilde c}_{\alpha}^*=u{\maru d}{}_{\alpha}^*+v^*s_{\alpha}{\maru d}{}_{\bar \alpha}\ , 
\nonumber\\
& &{\maru d}{}_{\bar \alpha}^*=u{\tilde c}_{\bar \alpha}^*+v^*s_{\alpha}{\tilde c}_{\alpha}\ , 
\quad\ {\rm i.e.,}\quad
{\tilde c}_{\bar \alpha}^*=u{\maru d}{}_{\bar \alpha}^*-v^*s_{\alpha}{\maru d}{}_{\alpha}\ . 
\label{2-6b}
\eeq
\esub
The operators $({\maru d}_\alpha, {\maru d}{}_{\alpha}^*)$ and 
$({\maru d}{}_{\bar \alpha}, {\maru d}{}_{\bar \alpha}^*)$ are called 
the quasiparticles, the use of which characterizes the approach B. 
The expectation values of ${\wtilde S}_{\pm,0}$ for $\rket{\phi}$ are given in the form 
\beq\label{2-7}
S_+=2\Omega_0uv^*\ , \qquad
S_-=2\Omega_0uv\ , \qquad
S_0=\Omega_0(|v|^2-u^2)\ . 
\eeq
With the use of the quasiparticle, the fluctuations around the values 
$S_{\pm,0}$ can be described, but, it is not the aim of this paper 
to present how to describe the fluctuations.

The orthogonal set is obtained by making successive operation of 
$({\maru d}{}_{\alpha}^*, {\maru d}{}_{\bar \alpha}^*)$ on the state $\rket{\phi}$:
\beq\label{2-8}
& &\rket{(\alpha),({\bar \alpha})}=
\prod_{i=1}^k {\maru d}{}_{\alpha_i}^*\prod_{j=1}^l{\maru d}{}_{{\bar \alpha}_j}^*
\rket{\phi}\ , \quad
\left({\wtilde D}_B^*=\prod_{i=1}^k {\maru d}{}_{\alpha_i}^*
\prod_{j=1}^l{\maru d}{}_{{\bar \alpha}_j}^*
\right)
\nonumber\\
& &k+l=n\ .
\eeq
The physical meaning of the state $\rket{(\alpha),({\bar \alpha})}$ can be interpreted 
in terms of the seniority number, that is, the number of the quasiparticles 
$n$. 
For example, ${\maru d}{}_{\alpha}^*\rket{\phi}$ (${\maru d}{}_{\bar \alpha}^*\rket{\phi}$) 
is the state with $n=1$ and two-quasiparticle states are the states with $n=2$. 
However, the two-quasiparticle state $\sum_{\alpha}s_{\alpha}{\maru d}{}_{\alpha}^*{\maru d}{}_{\bar \alpha}^*
\rket{\phi}$ should not be attributed to the state $n=2$ and certain consideration 
may be necessary, because this state is closely related to non-conservation of 
the fermion number.

As was mentioned in \S 1, we know the conventional algebraic approach, which seems to be 
different from the approach B. 
We call it the approach A in this paper. 
The first task of A is to investigate the structure of the minimum weight state $\rket{m}$ which 
does not contain a Cooper-pair. 
It obeys the condition 
\beq\label{2-9}
{\wtilde S}_-\rket{m}=0\ , \qquad 
{\wtilde S}_0\rket{m}=-s\rket{m}\ . \qquad
(s=0,\ 1/2,\ 1, \cdots )
\eeq
Then, the successive operation of ${\wtilde S}_+$ on the state $\rket{m}$ 
leads to the states composing the orthogonal basis. 
In this sense, the determination of $\rket{m}$ is unavoidable for the 
approach A. 
As a possible attempt, in (I), the present authors proposed an idea how to 
search $\rket{m}$. 
The basic idea of (I) can be found in a new $su(2)$-algebra 
$({\wtilde R}_{\pm,0})$, which satisfies 
\beq
& &[\ {\wtilde R}_+\ , \ {\wtilde R}_-\ ]=2{\wtilde R}_0\ , \qquad
[\ {\wtilde R}_0\ , \ {\wtilde R}_\pm\ ]=\pm {\wtilde R}_{\pm}\ , 
\label{2-10}\\
& &[\ {\rm any\ of}\ {\wtilde R}_{\pm,0}\ , \ {\rm any\ of}\ {\wtilde S}_{\pm,0}\ ]
=0\ . 
\label{2-11}
\eeq
Concrete expression of $({\wtilde R}_{\pm,0})$ is as follows: 
\beq\label{2-12}
{\wtilde R}_+=\sum_{\alpha}{\tilde c}_{\alpha}^*{\tilde c}_{\bar \alpha}\ , \quad
{\wtilde R}_-=\sum_{\alpha}{\tilde c}_{\bar \alpha}^*{\tilde c}_{\alpha}\ , \quad
{\wtilde R}_0=\frac{1}{2}\sum_{\alpha}
({\tilde c}_{\alpha}^*{\tilde c}_{\alpha}-{\tilde c}_{\bar \alpha}^*
{\tilde c}_{\bar \alpha})\ .
\eeq
In (I), we called $({\wtilde R}_{\pm,0})$ as the auxiliary algebra and the 
$R$-spin. 
The algebra $({\wtilde S}_{\pm,0})$, which we called the $S$-spin, 
plays a central role in the present model, because it describes the dynamics. 
The reason why we called the $R$-spin as the auxiliary algebra can be 
interpreted in the manner that the $R$-spin plays a central role 
for searching the minimum weight state. 
In this paper, we will attach the index $c$ to various quantities on which 
it was not necessary to attach the index $c$ in (I).

First, we will show the minimum weight state of the $R$-spin $\rket{m_0}$, 
which should be also the minimum weight state of the $S$-spin. 
According to (I), $\rket{m_0}$ can be expressed in the form composed of two parts, 
the aligned scheme and the $R$-spin scalar part: 
\beq\label{2-13}
\rket{m_0}={\wtilde {\cal P}}_c^*(p;(\lambda),q;(\mu\nu)){\wtilde {\cal Q}}_c^*(r;({\bar \alpha}))
\rket{0}\ . 
\eeq
In (I), we used the notation $\rket{r;({\bar \alpha})}$ for 
${\wtilde {\cal Q}}_c^*(r;({\bar \alpha}))\rket{0}$ in the expression 
\beq\label{2-14}
\rket{r;({\bar \alpha})}={\wtilde {\cal Q}}_c^*(r;({\bar \alpha}))\rket{0}\ , \qquad
{\wtilde {\cal Q}}_c^*(r;({\bar \alpha}))=\prod_{i=1}^{2r}{\tilde c}_{{\bar \alpha}_i}^*
\ . \quad (r=0,\ 1/2,\ 1,\cdots )\quad
\eeq
The state $\rket{r;({\bar \alpha})}$ belongs to the aligned scheme part and satisfies 
\beq\label{2-15}
{\wtilde R}_-\rket{r;({\bar \alpha})}=0\ , \qquad
{\wtilde R}_0\rket{r;({\bar \alpha})}=-r\rket{r;({\bar \alpha})}\ , \qquad
{\wtilde S}_-\rket{r;({\bar \alpha})}=0 \ . 
\eeq
The operator ${\wtilde {\cal P}}_c^*(p;(\lambda),q;(\mu\nu))$ is scalar for the 
$R$-spin and it composes the scalar part. 
The explicit form is as follows: 
\beq\label{2-16}
& &{\wtilde {\cal P}}_c^*(p;(\lambda),q;(\mu\nu))
=\prod_{k=1}^p{\wtilde {\cal S}}_{\lambda_k}^*(c)\prod_{j=1}^q{\wtilde {\cal S}}_{\mu_j\nu_j}^*(c)
\ , 
\eeq
\vspace{-0.2cm}
\bsub\label{2-17}
\beq
& &{\wtilde {\cal S}}_{\lambda}^*(c)
=\frac{1}{2}{\wtilde {\cal S}}_{\alpha\beta}^*(c)\ , \qquad (\alpha=\beta=\lambda)
\label{2-17a}\\
& &{\wtilde {\cal S}}_{\mu\nu}^*(c)={\wtilde {\cal S}}_{\alpha\beta}^*(c)\ . 
\qquad (\alpha=\mu \ , \ \beta=\nu\ , \ \mu\neq \nu)
\label{2-17b}
\eeq
\esub
Here, ${\wtilde {\cal S}}_{\alpha\beta}^*(c)$ is defined as 
\beq
& &{\wtilde {\cal S}}_{\alpha\beta}^*(c)
={\wtilde {\cal C}}_{\alpha}^*{\wtilde {\cal C}}_{\bar \beta}^*
-{\wtilde {\cal C}}_{\bar \alpha}^*{\wtilde {\cal C}}_{\beta}^* \ , 
\label{2-18}\\
& &
{\wtilde {\cal C}}_{\alpha}^*=\frac{1}{\Omega_0}
\left(-{\tilde c}_\alpha^*{\wtilde S}_0-\frac{1}{2}s_\alpha{\tilde c}_{\bar \alpha}{\wtilde S}_+
\right)\ , \quad
{\wtilde {\cal C}}_{\bar \alpha}^*=\frac{1}{\Omega_0}
\left(-{\tilde c}_{\bar \alpha}^*{\wtilde S}_0
+\frac{1}{2}s_\alpha{\tilde c}_{\alpha}{\wtilde S}_+
\right)\ .
\label{2-19} 
\eeq
The $R$-spin scalar operator ${\wtilde {\cal S}}_{\alpha\beta}^*(c)$ can be explicitly written as 
\beq\label{2-20}
{\wtilde{\cal S}}_{\alpha\beta}^*(c)
&=&
\frac{1}{\Omega_0^2}\biggl[
({\tilde c}_{\alpha}^*{\tilde c}_{\bar \beta}^*-{\tilde c}_{\bar \alpha}^*{\tilde c}_{\beta}^*)
{\wtilde S}_0\left({\wtilde S}_0-\frac{1}{2}\right)\nonumber\\
& &-\frac{1}{2}{\wtilde S}_+\left({\wtilde S}_0-\frac{1}{2}\right)
(s_{\beta}({\tilde c}_{\alpha}^*{\tilde c}_{\beta}
+{\tilde c}_{\bar \alpha}^*{\tilde c}_{\bar \beta}-\delta_{\alpha\beta})
+s_{\alpha}({\tilde c}_{\beta}^*{\tilde c}_{\alpha}
+{\tilde c}_{\bar \beta}^*{\tilde c}_{\bar \alpha}-\delta_{\alpha\beta}))\nonumber\\
& &-\frac{1}{4}\left({\wtilde S}_+\right)^2s_{\alpha}s_\beta
({\tilde c}_{\bar \beta}{\tilde c}_{\alpha}-{\tilde c}_{\beta}{\tilde c}_{\bar \alpha})\biggl]\ . 
\eeq
The factors $1/\Omega_0$ and $1/\Omega_0^2$ in the expressions (\ref{2-19}) and (\ref{2-20}) 
have been omitted in (I). 
But, the results including the conclusion are unchanged. 
Only some numerical factors connected to $1/\Omega_0$ and 
$1/\Omega_0^2$ appear. 
Since ${\wtilde {\cal S}}_{\alpha\beta}^*(c)$ is scalar for the $R$-spin, 
${\wtilde {\cal P}}_c^*(p;(\lambda),q;(\mu\nu))$ is also scalar. 
Then, $\rket{m}$ is given in a form 
\beq\label{2-21}
\rket{m}&=&
\left({\wtilde R}_+\right)^{r+r_0}\rket{m_0}
=\left({\wtilde R}_+\right)^{r+r_0}
{\wtilde {\cal P}}_c^*(p;(\lambda),q;(\mu\nu)){\wtilde {\cal Q}}_c^*(r;({\bar \alpha}))\rket{0}
\nonumber\\
&=&{\wtilde {\cal P}}_c^*(p;(\lambda),q;(\mu\nu))\left({\wtilde R}_+\right)^{r+r_0}
{\wtilde {\cal Q}}_c^*(r;({\bar \alpha}))\rket{0}\ . \nonumber\\
& &\left(
{\wtilde D}_A^*={\wtilde {\cal P}}_c^*(p;(\lambda),q;(\mu\nu))\left({\wtilde R}_+\right)^{r+r_0}
{\wtilde {\cal Q}}_c^*(r;({\bar \alpha}))\right)
\eeq
The state (\ref{2-21}) satisfies the relation (\ref{2-9}) with the condition 
\beq\label{2-22}
s=\Omega_0-(p+q+r)\ . 
\eeq
The state (\ref{2-21}) is also an eigenstate of ${\wtilde N}$ and 
its eigenvalue $n$ is given by 
\beq\label{2-23}
n=2(p+q+r)\ .
\eeq
Therefore, the seniority number is equal to $n=2(p+q+r)$, that is, 
the number of the fermions in the 
minimum weight state.

Thus, the eigenstate of ${\wtilde {\mib S}}^2$ and ${\wtilde S}_0$ with 
the eigenvalues $s(s+1)$ and $s_0$ is obtained in the form 
\beq\label{2-24}
\rdket{ss_0:p;(\lambda),q;(\mu\nu),rr_0;({\bar \alpha})}
&=&
\left({\wtilde S}_+\right)^{s+s_0}\rket{m}\nonumber\\
&=&
\left({\wtilde S}_+\right)^{s+s_0}{\wtilde {\cal P}}_c^*(p;(\lambda),q;(\mu\nu))
\left({\wtilde R}_+\right)^{r+r_0}{\wtilde {\cal Q}}_c^*(r;({\bar \alpha}))\rket{0}\ .
\nonumber\\
& &\left({\tilde C}_S^*=\left({\wtilde S}_+\right)^{s+s_0}\right)
\eeq
For the state (\ref{2-24}), the condition (\ref{2-22}) should be noted. 
Concerning the restriction for $(\lambda)$, $(\mu\nu)$ and $({\bar \alpha})$ and 
the orthogonality of the states (\ref{2-24}), we have discussed in detail in (I). 
Therefore, we will give only a short summary. 
The subscripts of any ${\wtilde {\cal S}}_{\mu_j\nu_j}^*(c)$ $(j=1,2,\cdots ,q)$ are 
all different from any $\alpha_i$ $(i=1,2,\cdots ,2r)$ related to 
${\wtilde {\cal Q}}_c^*(r;({\bar \alpha}))$ and on the occasion of successive operation of 
${\wtilde {\cal S}}_{\mu_j\nu_j}^*(c)$, 
all the subscripts are pairwise distinct. 
Concerning the orthogonality, generally, it may be impossible to prove 
if the states (\ref{2-21}) form an orthogonal basis or not. 
If not, we must derive an appropriate orthogonal basis from the set (\ref{2-21}). 
The above is the recapitulation of (I) based on the approach A.

\section{Quasiparticle in the framework of the conservation of the fermion number}

In the last section, we learned that the approaches A and B are in 
symmetrical relation. 
The approach B is formulated in the framework of non-conservation of the fermion number 
and the seniority coupling scheme is treated by operating the 
quasiparticles on the condensed Cooper-pair state. 
On the other hand, the approach A is formulated in the framework of the conservation of the fermion number 
and the Cooper-pairs are 
operated on the minimum weight state which describes the seniority coupling scheme. 
Therefore, it may be interesting to investigate if both approaches can be 
connected through the medium of a possible idea. 
As was mentioned in \S 1, the aim of this paper is to present 
the idea which will be called the approach C in this paper.

Instead of the condensed Cooper-pair state (\ref{2-4}), the approach C 
starts in the following Cooper-pair state: 
\beq\label{3-1}
\rket{\phi_\sigma}=\left({\wtilde S}_+\right)^{\sigma}\rket{0}\ . \qquad
(\sigma=0,\ 1,\ 2,\ \cdots ,\ 2\Omega_0)
\eeq
Clearly, the state $\rket{\phi_{\sigma}}$ is an eigenstate of ${\wtilde N}$ with 
the eigenvalue $2\sigma$ and we can prove the relation 
\beq\label{3-2}
{\tilde d}_{\alpha +}\rket{\phi_{\sigma}}={\tilde d}_{\alpha -}\rket{\phi_{\sigma}}
={\tilde d}_{{\bar \alpha} +}\rket{\phi_{\sigma}}
={\tilde d}_{{\bar \alpha} -}\rket{\phi_{\sigma}}=0\ . 
\eeq
The hermitian conjugate of ${\tilde d}_{\alpha +}$ etc. are expressed in the form 
\beq\label{3-3}
& &
{\tilde d}_{\alpha +}^*=\frac{1}{2\Omega_0}
\left[
(\Omega_0-{\wtilde S}_0){\tilde {c}}_{\alpha}^*
-s_{\alpha}{\tilde c}_{\bar \alpha}{\wtilde S}_+\right] , \ \ 
{\tilde d}_{\alpha -}^*=\frac{1}{2\Omega_0}
\left[
(\Omega_0+{\wtilde S}_0)s_{\alpha}{\tilde c}_{\bar \alpha}
-{\tilde {c}}_{\alpha}^*{\wtilde S}_-\right] ,
\nonumber\\
& &
{\tilde d}_{{\bar \alpha} +}^*=\frac{1}{2\Omega_0}
\left[
(\Omega_0-{\wtilde S}_0){\tilde {c}}_{\bar \alpha}^*
+s_{\alpha}{\tilde c}_{\alpha}{\wtilde S}_+\right] , \ \ 
{\tilde d}_{{\bar \alpha} -}^*=\frac{1}{2\Omega_0}
\left[
-(\Omega_0+{\wtilde S}_0)s_{\alpha}{\tilde c}_{\alpha}
-{\tilde {c}}_{\bar \alpha}^*{\wtilde S}_-\right] . \qquad
\eeq
The relation (\ref{3-2}) corresponds to the relation (\ref{2-6a}). 
The sets $({\tilde d}_{\alpha +}^*, {\tilde d}_{\alpha -}^*)$ 
and $({\tilde d}_{{\bar \alpha} +}^*, {\tilde d}_{{\bar \alpha} -}^*)$ form 
the $S$-spin spinors, respectively: 
\bsub\label{3-4}
\beq
& &[\ {\tilde S}_{\mp}\ , \ {\tilde d}_{\alpha \pm}^*\ ]
={\tilde d}_{\alpha \mp}^*\ , \qquad
[\ {\tilde S}_{0}\ , \ {\tilde d}_{\alpha \pm}^*\ ]
=\pm\frac{1}{2}{\tilde d}_{\alpha \pm}^*\ , \qquad
[\ {\tilde S}_{\pm}\ , \ {\tilde d}_{\alpha \pm}^*\ ]
=0\ , \nonumber\\
& &[\ {\tilde S}_{\mp}\ , \ {\tilde d}_{{\bar \alpha} \pm}^*\ ]
={\tilde d}_{{\bar \alpha} \mp}^*\ , \qquad
[\ {\tilde S}_{0}\ , \ {\tilde d}_{{\bar \alpha} \pm}^*\ ]
=\pm\frac{1}{2}{\tilde d}_{{\bar \alpha} \pm}^*\ , \qquad
[\ {\tilde S}_{\pm}\ , \ {\tilde d}_{{\bar \alpha} \pm}^*\ ]
=0\ .
\label{3-4a}
\eeq
The sets $({\tilde d}_{\alpha +}^*, {\tilde d}_{{\bar \alpha} +}^*)$ and 
$({\tilde d}_{\alpha -}^*, {\tilde d}_{{\bar \alpha} -}^*)$, 
the combinations of which are different of the $S$-spin spinors, form the 
$R$-spin spinors, respectively: 
\beq
& &[\ {\tilde R}_{-}\ , \ {\tilde d}_{\alpha \pm}^*\ ]
={\tilde d}_{{\bar \alpha} \pm}^*\ , \qquad
[\ {\tilde R}_{0}\ , \ {\tilde d}_{\alpha \pm}^*\ ]
=\frac{1}{2}{\tilde d}_{\alpha \pm}^*\ , \qquad
[\ {\tilde R}_{+}\ , \ {\tilde d}_{\alpha \pm}^*\ ]
=0\ , \nonumber\\
& &[\ {\tilde R}_{+}\ , \ {\tilde d}_{{\bar \alpha} \pm}^*\ ]
={\tilde d}_{{\alpha} \pm}^*\ , \qquad
[\ {\tilde R}_{0}\ , \ {\tilde d}_{{\bar \alpha} \pm}^*\ ]
=-\frac{1}{2}{\tilde d}_{{\bar \alpha} \pm}^*\ , \qquad
[\ {\tilde R}_{-}\ , \ {\tilde d}_{{\bar \alpha} \pm}^*\ ]
=0\ .\quad
\label{3-4b}
\eeq
\esub
The case of ${\tilde d}_{\alpha +}$ etc. is obtained in terms of the hermitian conjugate 
of the relation (\ref{3-4}). 

The inverse of the form (\ref{3-3}) will be given after some discussion. 
Our present fermion-space is constructed in terms of the four operators, 
${\tilde c}_{\alpha}^*$, ${\tilde c}_{\bar \alpha}^*$, ${\tilde c}_{\alpha}$ and ${\tilde c}_{\bar \alpha}$ 
for the partner $(\alpha, {\bar \alpha})$. 
However, we intend to describe the present model 
in terms of the eight operators,  
${\tilde d}_{\alpha \pm}^*$, ${\tilde d}_{{\bar \alpha} \pm}^*$, ${\tilde d}_{\alpha \pm}$ and ${\tilde d}_{{\bar \alpha}\pm}$. 
Therefore, there should exist four constraints for the above eight operators and it is enough to adopt four 
operators for obtaining the inverse of the relation (\ref{3-3}). 
After rather lengthy consideration for the relation (\ref{3-3}) and its hermitian conjugate, we have the following form: 
\bsub\label{3-5}
\beq
& &{\tilde c}_{\alpha}^*={\wtilde Z}_0\left[(\Omega_0-{\wtilde S}_0+1){\tilde d}_{\alpha +}^*+{\wtilde S}_+s_{\alpha}{\tilde d}_{{\bar \alpha} +}\right] \ , 
\label{3-5a}\\
& &{\tilde c}_{\bar \alpha}^*={\wtilde Z}_0\left[(\Omega_0-{\wtilde S}_0+1){\tilde d}_{{\bar \alpha} +}^*-{\wtilde S}_+s_{\alpha}{\tilde d}_{\alpha +}\right] \ . 
\label{3-5b}
\eeq
\esub 
Here, ${\wtilde Z}_0$ is defined as 
\beq\label{3-6}
{\wtilde Z}_0=2\Omega_0\left[{\wtilde S}_+{\wtilde S}_-+\left(\Omega_0-{\wtilde S}_0+\frac{1}{2}\right)
(\Omega_0-{\wtilde S}_0+1)\right]^{-1}\ .
\eeq
Since ${\wtilde S}_+{\wtilde S}_-$ is positive-definite and $s_0$, the eigenvalue of ${\wtilde S}_0$, obeys 
the condition $-\Omega_0\leq s_0 \leq \Omega_0$, the fractional operator ${\wtilde Z}_0$ is definable. 
Further, we have 
\bsub\label{3-7}
\beq
& &{\tilde d}_{\alpha -}^*=-{\tilde d}_{\alpha +}^*{\wtilde S}_-\left(\frac{4\Omega_0+1}{4\Omega_0}\right){\wtilde Z}_0
-s_{\alpha}{\tilde d}_{{\bar \alpha}+}\left[1-(\Omega_0-{\wtilde S}_0+1)
\left(\frac{4\Omega_0+1}{4\Omega_0}\right){\wtilde Z}_0\right]\ , \nonumber\\
& &\label{3-7a}\\
& &{\tilde d}_{{\bar \alpha} -}^*=-{\tilde d}_{{\bar \alpha} +}^*{\wtilde S}_-\left(\frac{4\Omega_0+1}{4\Omega_0}\right){\wtilde Z}_0
+s_{\alpha}{\tilde d}_{{\alpha}+}\left[1-(\Omega_0-{\wtilde S}_0+1)
\left(\frac{4\Omega_0+1}{4\Omega_0}\right){\wtilde Z}_0\right]\ . \nonumber\\
& &\label{3-7b}
\eeq
\esub
Of course, the hermitian conjugate of the relations (\ref{3-5}) and (\ref{3-7}) are also available. 
The relation (\ref{3-5}) tells us that $({\tilde c}_{\alpha}^*, {\tilde c}_{{\bar \alpha}}^*, {\tilde c}_{\alpha}, {\tilde c}_{\bar \alpha})$ 
are expressed as linear functions of 
$({\tilde d}_{\alpha +}^*, {\tilde d}_{{\bar \alpha}+}^*, {\tilde d}_{\alpha +}, {\tilde d}_{{\bar \alpha}+})$ with the ${\wtilde S}_{\pm, 0}$-dependent coefficients. 
The relation (\ref{3-7}) shows that 
$({\tilde d}_{\alpha -}^*, {\tilde d}_{{\bar \alpha}-}^*, {\tilde d}_{\alpha -}, {\tilde d}_{{\bar \alpha}-})$ are functions of 
$({\tilde d}_{\alpha +}^*, {\tilde d}_{{\bar \alpha}+}^*, {\tilde d}_{\alpha +}, {\tilde d}_{{\bar \alpha}+})$, namely, 
the four constraints. 
However, for treating various relations for the present model, it may be convenient to treat the eight operators impartially under the 
constraints. 
The above is the inverse of the relation (\ref{3-3}) combined under the constraints.

As is well known, the approach B is based on the mean field approximation, 
which suggests that, as the zeroth order approximation, 
${\wtilde S}_{\pm,0}$ can be replaced with the $c$-numbers 
$S_{\pm,0}$ shown in the relation (\ref{2-7}). 
We call this procedure as the $c$-number replacement. 
Then, it may be interesting to investigate various relations presented in this section 
under the $c$-number replacement. 
In this procedure, we must note the following: 
For example, we consider the case ${\wtilde S}_+s_{\alpha}{\tilde c}_{\bar \alpha}$, which can be 
rewritten to $(s_{\alpha}{\tilde c}_{\bar \alpha}{\wtilde S}_++{\tilde c}_{\alpha}^*)$. 
Last term ${\tilde c}_{\alpha}^*$ indicates the fluctuation around $S_+s_{\alpha}{\tilde c}_{\bar \alpha}$. 
Therefore, if we follow the basic viewpoint of the mean field approximation, the $c$-number replacement 
is performed in the way 
\beq\label{3-8}
{\wtilde S}_+s_{\alpha}{\tilde c}_{\bar \alpha}\ , \quad s_{\alpha}{\tilde c}_{\bar \alpha}{\wtilde S}_+
\longrightarrow 
S_+s_{\alpha}{\tilde c}_{\bar \alpha}\ .
\eeq 
The above is an example of the $c$-number replacement.

Following the idea of the $c$-number replacement, for example, such as the replacement (\ref{3-8}), 
let us replace ${\wtilde S}_{\pm,0}$ contained in the relation (\ref{3-3}) with $S_{\pm,0}$. 
After the replacement, we have the following: 
\beq\label{3-9}
& &{\tilde d}_{\alpha +}^*\longrightarrow u{\maru d}{}^*_{\alpha}\ , \qquad 
{\tilde d}_{\alpha -}^*\longrightarrow -v{\maru d}{}^*_{\alpha}\ , \nonumber\\
& &{\tilde d}_{{\bar \alpha} +}^*\longrightarrow u{\maru d}{}^*_{\bar \alpha}\ ,\qquad
{\tilde d}_{{\bar \alpha} -}^*\longrightarrow -v{\maru d}{}^*_{\bar \alpha}\ .
\eeq
For example, the case ${\maru d}{}_{\alpha +}^*$ is derived through the process 
\beq\label{3-10}
{\tilde d}_{\alpha +}^*\longrightarrow & &
\frac{1}{2\Omega_0}\left[(\Omega_0-S_0){\tilde c}_{\alpha}^*-S_+s_{\alpha}{\tilde c}_{\bar \alpha}\right]
=u^2{\tilde c}_{\alpha}^*-uv^*s_{\alpha}{\tilde c}_{\bar \alpha} \nonumber\\
& &=u(u{\tilde c}_{\alpha}^*-v^*s_{\alpha}{\tilde c}_{\bar \alpha})=u{\maru d}{}_{\alpha}^* \ .
\eeq
As can be seen in the relation (\ref{3-3}), the fermion number increases by 1 under 
one-time operation of ${\tilde d}_{\alpha +}^*$ (${\tilde d}_{{\bar \alpha}+}^*$) and 
decreases by 1 under one-time operation of ${\tilde d}_{\alpha -}^*$ (${\tilde d}_{{\bar \alpha}-}^*$). 
However, the relation (\ref{3-9}) tells us that in the approach B, there does not exist any distinction between 
${\tilde d}_{\alpha +}^*$ (${\tilde d}_{{\bar \alpha}+}^*$) and ${\tilde d}_{\alpha -}^*$ (${\tilde d}_{{\bar \alpha}-}^*$) 
except the $c$-numbers $u$ and $v$. 
It comes from the difference between the state $\rket{\phi}$ and $\rket{\phi_{\sigma}}$, 
non-conservation and conservation of the fermion number.  
Concerning the relations (\ref{3-5}) and (\ref{3-7}), the following replacements are added: 
\beq\label{3-11}
4\Omega_0+1 \longrightarrow 4\Omega_0\ , \quad
\Omega_0-{\wtilde S}_0+\frac{1}{2}\ \ {\rm and}\ \ \Omega_0-{\wtilde S}_0+1\longrightarrow \Omega_0-S_0=2\Omega_0u^2\ .\quad\ \ 
\eeq
Here, the $c$-numbers 1 and 1/2 may be regarded as the fluctuations around $4\Omega_0$ and 
$\Omega_0-S_0$, respectively. 
Under the above replacement, ${\wtilde Z}_0$ is reduced to the following:
\beq\label{3-12}
{\wtilde Z}_0\longrightarrow 2\Omega_0(4\Omega_0^2u^2|v|^2+4\Omega_0^2u^4)^{-1}=
\frac{1}{2\Omega_0u^2}\ .
\eeq
Then, the relation (\ref{3-5}) gives 
\beq\label{3-13}
{\tilde c}_{\alpha}^*\longrightarrow u{\maru d}{}_{\alpha}^*+v^*s_{\alpha}{\maru d}{}_{\bar \alpha}\ , \qquad
{\tilde c}_{\bar \alpha}^*\longrightarrow u{\maru d}{}_{\bar \alpha}^*-v^*s_{\alpha}{\maru d}{}_{\alpha}\ .
\eeq
The above is nothing but the relation (\ref{2-6b}). 
Concerning the relation (\ref{3-7}), both sides become identical. 
For example, the case (\ref{3-7a}) is the following: 
The left-hand side is reduced to $-v{\maru d}{}_{\alpha}^*$, 
which is shown in the relation (\ref{3-9}). 
On the other hand, the right-hand side becomes the following: 
\beq\label{3-14}
& &\hbox{\rm the\ right-hand\ side}\longrightarrow \nonumber\\
& &-u{\maru d}{}_{\alpha}^*\cdot 2\Omega_0 uv\cdot \frac{4\Omega_0}{4\Omega_0}\cdot
\frac{1}{2\Omega_0u^2}
-us_{\alpha}{\maru d}{}_{\bar \alpha}^*\left[
1-2\Omega_0u^2\cdot\frac{4\Omega_0}{4\Omega_0}\cdot\frac{1}{2\Omega_0u^2}\right]
=-v{\maru d}{}_{\alpha}^*\ . \quad\ 
\eeq
Certainly, the both sides become equally $-v{\maru d}{}_{\alpha}^*$.

We can learn from the above investigation that the operator ${\tilde d}_{\alpha +}^*$ etc. 
introduced in this section may be regarded as a possible variety of the quasiparticles 
${\maru d}{}_{\alpha}^*$ etc. to the case of the framework of the conservation of the fermion number. 
Of course, these new operators are not fermion operators and, then, the handling of these operators is not so 
easy as that of the quasiparticles. 
But, the role is very similar to that of the quasiparticles. 
In this sense, we call the new operator as quasiparticle in the framework of the conservation of the fermion number 
and abbreviate it as the ``quasiparticle."
As a final comment, we mention the spinor property of ${\maru d}_{\alpha}$, ${\maru d}{}_{\alpha}^*$, 
${\maru d}_{\bar \alpha}$ and ${\maru d}{}_{\bar \alpha}^*$. 
In the relations (\ref{3-4a}) and (\ref{3-4b}), we showed the spinor property of 
${\maru d}{}_{\alpha\pm}^*$ and ${\maru d}{}_{{\bar \alpha}\pm}^*$ for the $S$- and the $R$-spin. 
However, with the use of the relation which 
will be shown in the relation (\ref{5-17}), we can prove that 
$({\maru d}{}_{\alpha}^*$, ${\maru d}{}_{{\bar \alpha}}^*)$ forms the $R$-spin spinor, but 
any combination of ${\maru d}_{\alpha}$, ${\maru d}{}_{\alpha}^*$, 
${\maru d}_{\bar \alpha}$ and ${\maru d}{}_{\bar \alpha}^*$ does not form the 
$S$-spin spinor. 
The above is essential difference between the ``quasiparticle" and the 
quasiparticle. 
The spinor property of 
${\tilde d}_{\alpha\pm}^*$ and ${\tilde d}_{{\bar \alpha}\pm}^*$ for both spins 
will play a basic role in the discussions of \S\S 4 and 5.

\section{The minimum weight state in terms of the ``quasiparticle"}

Main aim of this section is to express the minimum weight state $\rket{m}$ shown in the relation (\ref{2-21}) 
in terms of the ``quasiparticle." 
First, we express the aligned scheme part $\rket{r;({\bar \alpha})}$ shown in the relation 
(\ref{2-14}). 
Direct calculation leads us to the following: 
\beq\label{4-1}
& &\rket{r;({\bar \alpha})}={\wtilde {\cal Q}}_c^*(r;({\bar \alpha}))\rket{0}
=\frac{(4\Omega_0-2r-1)!(4\Omega_0)^{2r}}{(4\Omega_0-1)!}{\wtilde {\cal Q}}_d^*(r;({\bar \alpha}))\rket{0} \ , 
\nonumber\\
& &{\wtilde {\cal Q}}_c^*(r;({\bar \alpha}))=\prod_{i=1}^{2r}{\tilde c}_{{\bar \alpha}_i}^*\ , \qquad
{\wtilde {\cal Q}}_d^*(r;({\bar \alpha}))=\prod_{i=1}^{2r}{\tilde d}_{{\bar \alpha}_i+}^*\ . 
\eeq
We can see that the aligned scheme part is expressed as a function of the 
``quasiparticle" and the functional form is the same as that of the original expression 
except the numerical factor. 
The form (\ref{4-1}) may be easily understood by the definition of $\rket{r;({\bar \alpha})}$ and 
the relation (\ref{3-3}).
The second term of the right-hand of ${\tilde c}_{{\bar \alpha}_i}^*$ in (\ref{3-5b}) does not 
give any influence. 
The relation (\ref{3-4b}) tells us that the operation of ${\wtilde R}_+$ on the state 
$\prod_{i=1}^{2r}{\tilde d}_{{\bar \alpha}_i+}^*\rket{0}$ gives the same result of the case 
$\prod_{i=1}^{2r}{\tilde c}_{{\bar \alpha}_i}^*\rket{0}$.

Next, we will investigate the scalar part. 
Before entering the central part, we repeat a preliminary argument which has been already performed in (I). 
Let us define an operator ${\wtilde {\cal P}}^*$, which is associated with ${\wtilde P}^*$, in the form 
\beq\label{4-2}
{\wtilde {\cal P}}^*=\frac{1}{\Omega_0}\left(-{\wtilde P}^*{\wtilde S}_0
-\frac{1}{2}[\ {\wtilde S}_-\ , \ {\wtilde P}^*\ ]{\wtilde S}_+\right)\ . 
\eeq
The operator ${\wtilde {\cal P}}^*$ satisfies the relation 
\beq\label{4-3}
[\ {\wtilde S}_-\ , \ {\wtilde {\cal P}}^*\ ]=-\frac{1}{\Omega_0}{\wtilde P}^*{\wtilde S}_-\ , \quad {\rm i.e.,}\quad 
{\wtilde S}_-{\wtilde {\cal P}}^*=\left({\wtilde {\cal P}}^*-\frac{1}{\Omega_0}{\wtilde P}^*\right){\wtilde S}_-\ , 
\eeq
if ${\wtilde P}^*$ obeys the condition 
\beq\label{4-4}
[\ {\wtilde S}_-\ , \ [\ {\wtilde S}_-\ , \ {\wtilde P}^*\ ]\ ]=0\ .
\eeq
The factor $1/\Omega_0$ in the relations (\ref{4-2}) and (\ref{4-3}) have been omitted in (I). 
But, the results including the conclusion are unchanged. 
Only some numerical factors connected to $1/\Omega_0$ appear. 
We can learn that the operation of ${\wtilde S}_-$ from the left-side of ${\wtilde {\cal P}}^*$ is 
equivalent to the operation of ${\wtilde S}_-$ from the right-side of 
$({\wtilde {\cal P}}^*-{\wtilde P}^*/\Omega_0)$. 
An example of ${\wtilde P}^*$ is $({\tilde c}_{\alpha}^*, {\tilde c}_{\bar \alpha}^*)$. 
For example, we have 
\beq\label{4-5}
[\ {\wtilde S}_-\ , \ [\ {\wtilde S}_-\ , \ {\tilde c}_{\alpha}^*\ ]\ ]=
[\ {\wtilde S}_-\ , \ s_{\alpha}{\tilde c}_{\bar \alpha}\ ]=0\ .
\eeq
The property (\ref{4-3}) permits to use ${\wtilde {\cal C}}_{\alpha}^*$ and ${\wtilde {\cal C}}_{\bar \alpha}^*$ 
given in the form (\ref{2-19}) for the construction of $\rket{m}$ which obeys 
${\wtilde S}_-\rket{m}=0$. 
We apply the above idea to the operators ${\tilde d}_{\alpha +}^*$ and ${\tilde d}_{{\bar \alpha}+}^*$: 
\beq\label{4-6}
& &[\ {\wtilde S}_-\ , \ [\ {\wtilde S}_-\ , \ {\tilde d}_{\alpha +}^*\ ]\ ]=
[\ {\wtilde S}_-\ , \ {\tilde d}_{\alpha -}^*\ ]=0\ , \nonumber\\
& &[\ {\wtilde S}_-\ , \ [\ {\wtilde S}_-\ , \ {\tilde d}_{{\bar \alpha} +}^*\ ]\ ]=
[\ {\wtilde S}_-\ , \ {\tilde d}_{{\bar \alpha} -}^*\ ]=0\ . 
\eeq
For the above derivation, we used the relation (\ref{3-4a}). 
Then, we define the following operators 
\beq\label{4-7}
{\wtilde {\cal D}}_{\alpha}^*=\frac{1}{\Omega_0}
\left(-{\tilde d}_{\alpha +}^*{\wtilde S}_0-\frac{1}{2}{\tilde d}_{\alpha -}^*{\wtilde S}_+\right)\ , \quad
{\wtilde {\cal D}}_{\bar \alpha}^*=\frac{1}{\Omega_0}
\left(-{\tilde d}_{{\bar \alpha} +}^*{\wtilde S}_0-\frac{1}{2}{\tilde d}_{{\bar \alpha} -}^*{\wtilde S}_+\right)\ . 
\eeq
In the same idea as that shown in the relation (\ref{2-18}), we define 
\beq\label{4-8}
{\wtilde {\cal S}}_{\alpha\beta}^*(d)={\wtilde {\cal D}}_{\alpha}^*{\wtilde {\cal D}}_{\bar \beta}^*
-{\wtilde {\cal D}}_{\bar \alpha}^*{\wtilde {\cal D}}_{\beta}^*\ . 
\eeq
The relation (\ref{3-4b}) tells us that ${\wtilde {\cal S}}_{\alpha\beta}^*(d)$ is also the 
$R$-spin spinor. 
After straightforward calculation, we can derive the following relation: 
\beq
& &{\wtilde {\cal S}}_{\alpha\beta}^*(d)={\wtilde {\cal S}}_{\alpha\beta}^*(c)\cdot
\frac{1}{(2\Omega_0)^2}(\Omega_0-{\wtilde S}_0)
\left(\Omega_0-{\wtilde S}_0+\frac{1}{2}\right)+{\wtilde S}_+\cdot {\wtilde {\cal T}}_{\alpha\beta}^*\cdot {\wtilde S}_- \ , 
\label{4-9}\\
& &{\wtilde {\cal T}}_{\alpha\beta}^*=
\frac{4}{(2\Omega_0)^4}\biggl[
({\tilde c}_{\alpha}^*{\tilde c}_{\bar \beta}^*-{\tilde c}_{\bar \alpha}^*{\tilde c}_{\beta}^*)
\left(\frac{1}{4}{\wtilde S}_+{\wtilde S}_--\left({\wtilde S}_0+\frac{1}{2}\right)\left(\Omega_0-{\wtilde S}_0-\frac{1}{4}\right)\right)
\nonumber\\
& &\qquad\qquad\qquad
+\frac{1}{4}{\wtilde S}_+(s_{\beta}({\tilde c}_{\alpha}^*{\tilde c}_{\beta}+{\tilde c}_{\bar \alpha}^*{\tilde c}_{\bar \beta}-\delta_{\alpha\beta})
+s_{\alpha}({\tilde c}_{\beta}^*{\tilde c}_{\alpha}+{\tilde c}_{\bar \beta}^*{\tilde c}_{\bar \alpha}-\delta_{\alpha\beta}))
(\Omega_0-{\wtilde S}_0-1)
\nonumber\\
& &\qquad\qquad\qquad
-\frac{1}{8}{\wtilde S}_+s_{\beta}({\tilde c}_{\alpha}^*{\tilde c}_{\beta}+{\tilde c}_{\bar \alpha}^*{\tilde c}_{\bar \beta})\biggl] \ .
\label{4-10}
\eeq
The existence of ${\wtilde S}_-$ in the second term of 
${\wtilde {\cal S}}_{\alpha\beta}^*(d)$ teaches us that, in the minimum weight state $\rket{m}$, the second term vanishes. 
It suggests us that the role of ${\wtilde {\cal S}}_{\alpha\beta}^*(d)$ is essentially the same as that of 
${\wtilde {\cal S}}_{\alpha\beta}^*(c)$.

On the basis of the above argument, we will arrange how to construct the minimum weight state $\rket{m}$ in 
terms of the ``quasiparticle." 
Let $\rket{m}$ be expressed in the form 
\beq\label{4-11}
& &\rket{m}={\wtilde {\cal P}}_d^*(p;(\lambda),q;(\mu\nu))\left({\wtilde R}_+\right)^{r+r_0}{\wtilde {\cal Q}}_d^*(r;({\bar \alpha}))\rket{0}\ . 
\nonumber\\
& &\qquad \left({\wtilde D}_C^*={\wtilde {\cal P}}_d^*(p;(\lambda),q;(\mu\nu))\left({\wtilde R}_+\right)^{r+r_0}{\wtilde {\cal Q}}_d^*(r;({\bar \alpha}))
\right)
\eeq
Here, ${\wtilde {\cal Q}}_d^*(r;({\bar \alpha}))$ is given in the relation (\ref{4-1}). 
The operator ${\wtilde {\cal P}}_d^*$ is defined as follows: 
\beq\label{4-12}
{\wtilde {\cal P}}_d^*(p;(\lambda),q;(\mu\nu))=\prod_{k=1}^p{\wtilde {\cal S}}_{\lambda_k}^*(d)\prod_{j=1}^q{\wtilde {\cal S}}_{\mu_j\nu_j}^*(d) \ , 
\eeq
\vspace{-0.6cm}
\bsub\label{4-13}
\beq
& &{\wtilde {\cal S}}_{\lambda}^*(d)=\frac{1}{2}{\wtilde {\cal S}}_{\alpha\beta}^*(d)\ , \qquad (\alpha=\beta=\lambda)
\qquad\qquad
\label{4-13a}\\
& &{\wtilde {\cal S}}_{\mu\nu}^*(d)={\wtilde {\cal S}}_{\alpha\beta}^*(d)\ , \qquad (\alpha=\mu,\ \beta=\nu, \ \mu\neq \nu)
\label{4-13b}
\eeq
\esub
Here, ${\wtilde {\cal S}}_{\alpha\beta}^*(d)$ is given in the relation (\ref{4-8}). 
It may be self-evident that the state (\ref{4-11}) is the minimum weight state. 
By using the relations (\ref{4-1}) and (\ref{4-9}), we can connect the expression (\ref{4-11}) with the form 
(\ref{2-21}) in the following way: 
\beq\label{4-14}
& &
\rdket{p;(\lambda),q;(\mu\nu),rr_0;({\bar \alpha})}
\nonumber\\
&=&{\wtilde {\cal P}}_d^*(p;(\lambda),q;(\mu\nu))\left({\wtilde R}_+\right)^{r+r_0}
{\wtilde {\cal Q}}_d^*(r;({\bar \alpha}))\rket{0}
\nonumber\\
&=&\frac{1}{(4\Omega_0)^n}\cdot\frac{(4\Omega_0-1)!}{(4\Omega_0-n+1)!}
\cdot\frac{(4\Omega_0-2r+1)!}{(4\Omega_0-2r-1)!}
\nonumber\\
& &\times {\wtilde {\cal D}}_c^*(p;(\lambda),q;(\mu\nu))\left({\wtilde R}_+\right)^{r+r_0}
{\wtilde {\cal Q}}_c^*(r;({\bar \alpha}))\rket{0}\ . 
\quad
(n=2(p+q+r))
\eeq
We can see that both expressions are essentially equivalent. 
Therefore, instead of the state (\ref{2-24}), we adopt the form 
\beq\label{4-15}
& &
\rdket{ss_0:p;(\lambda),q;(\mu\nu),rr_0;({\bar \alpha})}
=\left({\wtilde S}_+\right)^{s+s_0}\rdket{p;(\lambda),q;(\mu\nu),rr_0;({\bar \alpha})}
\nonumber\\
&=&\left({\wtilde S}_+\right)^{s+s_0}
{\wtilde {\cal P}}_d^*(p;(\lambda),q;(\mu\nu))\left({\wtilde R}_+\right)^{r+r_0}
{\wtilde {\cal Q}}_d^*(r;({\bar \alpha}))\rket{0}\ .
\eeq
The state (\ref{4-15}) is the re-formation of the state (\ref{2-24}). 
Therefore, as was mentioned in the place after the state (\ref{2-24}), the restriction 
for $(\lambda)$, $(\mu\nu)$ and $({\bar \alpha})$ and the orthogonality of the state 
(\ref{4-15}) are in the same situations as those in the state (\ref{2-24}). 
In the next section, we will investigate the connection of the 
state (\ref{4-15}) to the approach B. 
The above is a part of the approach C.

\section{Re-formation of the approach C presented in \S 4}

As was already mentioned, the orthogonal set in the approach B is constructed 
by operating the quasiparticles appropriately on the condensed 
Cooper-pair state $\rket{\phi}$. 
On the other hand, the structure of the state (\ref{4-15}) is the reverse of the state (\ref{2-8}) in the approach B. 
The state (\ref{4-15}) is obtained by operating ${\wtilde S}_+$ appropriately on the 
minimum weight state $\rket{m}$ constructed by the ``quasiparticle." 
Therefore, in order to compare the set of the states obtained in \S 4 
with the orthogonal set of the approach B, we must 
reverse the order of the operations of ${\wtilde S}_+$ and the 
``quasiparticle" on $\rket{0}$, namely, the re-formation 
of the approach C.

For the above-mentioned aim, we consider the relation 
\bsub\label{5-1}
\beq
& &[\ {\wtilde S}_+\ , \ {\wtilde {\cal D}}_\alpha^*\ ]=\frac{1}{2\Omega_0}{\tilde d}_{\alpha +}^*{\wtilde S}_+\ , 
\quad
{\rm i.e.,}\quad
{\wtilde S}_+{\wtilde {\cal D}}_{\alpha}^*=\left({\wtilde {\cal D}}_{\alpha}^*+\frac{1}{2\Omega_0}{\tilde d}_{\alpha +}^*
\right){\wtilde S}_+\ , 
\label{5-1a}\\
& &[\ {\wtilde S}_+\ , \ {\wtilde {\cal D}}_{\bar \alpha}^*\ ]=\frac{1}{2\Omega_0}{\tilde d}_{{\bar \alpha} +}^*{\wtilde S}_+\ , 
\quad
{\rm i.e.,}\quad
{\wtilde S}_+{\wtilde {\cal D}}_{\bar \alpha}^*=\left({\wtilde {\cal D}}_{{\bar \alpha}}^*+\frac{1}{2\Omega_0}{\tilde d}_{{\bar \alpha} +}^*
\right){\wtilde S}_+\ . 
\label{5-1b}
\eeq
\esub
With the successive use of the relation (\ref{5-1}), we have 
\bsub\label{5-2}
\beq
& &\left({\wtilde S}_+\right)^{\sigma}{\wtilde {\cal D}}_{\alpha}^*
={\wtilde {\cal D}}_{\alpha,\sigma}^*\left({\wtilde S}_+\right)^{\sigma}\ , \qquad
{\wtilde {\cal D}}_{\alpha,\sigma}^*={\wtilde {\cal D}}_{\alpha}^*+\frac{\sigma}{2\Omega_0}{\tilde d}_{\alpha +}^*\ , 
\label{5-2a}\\
& &\left({\wtilde S}_+\right)^{\sigma}{\wtilde {\cal D}}_{{\bar \alpha}}^*
={\wtilde {\cal D}}_{{\bar \alpha},\sigma}^*\left({\wtilde S}_+\right)^{\sigma}\ , \qquad
{\wtilde {\cal D}}_{{\bar \alpha},\sigma}^*={\wtilde {\cal D}}_{{\bar \alpha}}^*+\frac{\sigma}{2\Omega_0}{\tilde d}_{{\bar \alpha} +}^*\ . 
\label{5-2b}
\eeq
\esub
Then, we can define a $R$-spin scalar in the following form:
\beq\label{5-3}
{\wtilde {\cal S}}_{\alpha\beta,\sigma}^*(d)
={\wtilde {\cal D}}_{\alpha,\sigma}^*{\wtilde {\cal D}}_{{\bar \beta},\sigma}^*
-{\wtilde {\cal D}}_{{\bar \alpha},\sigma}^*{\wtilde {\cal D}}_{{\beta},\sigma}^* \ .
\eeq
The operator ${\wtilde {\cal S}}_{\alpha\beta,\sigma}^*(d)$ obeys the relation 
\beq\label{5-4}
\left({\wtilde S}_+\right)^{\sigma}{\wtilde {\cal S}}_{\alpha\beta}^*(d)={\wtilde {\cal S}}_{\alpha\beta,\sigma}^*(d)
\left({\wtilde S}_+\right)^{\sigma}\ .
\eeq
Further, we have 
\beq
& &[\ {\wtilde S}_+\ , \ {\wtilde R}_+\ ]=0\ , \quad {\rm i.e.,}\quad
\left({\wtilde S}_+\right)^{\sigma}\left({\wtilde R}_+\right)^{r+r_0}
=\left({\wtilde R}_+\right)^{r+r_0}\left({\wtilde S}_+\right)^{\sigma}\ , 
\label{5-5}\\
& &[\ {\wtilde S}_+\ , \ {\tilde d}_{{\bar \alpha}+}^*\ ]=0\ , \quad {\rm i.e.,}\quad
\left({\wtilde S}_+\right)^{\sigma}{\tilde d}_{{\bar \alpha}+}^*
={\tilde d}_{{\bar \alpha}+}^*\left({\wtilde S}_+\right)^{\sigma}\ . 
\label{5-6}
\eeq
We can see that in the relations (\ref{5-4})-({\ref{5-6}), 
the operation of $({\wtilde S}_+)^{\sigma}$ from the left-side changes 
to that from the right-side. 
With the use of relations (\ref{5-4})-(\ref{5-6}), 
the state (\ref{4-15}) can be re-formed to 
\beq
& &\rdket{ss_0:p;(\lambda),q;(\mu\nu),rr_0;({\bar \alpha})}\nonumber\\
&=&{\wtilde {\cal P}}_{d,\sigma}^*(p;(\lambda),q;(\mu\nu))\left({\wtilde R}_+\right)^{r+r_0}
{\wtilde {\cal Q}}_{d}^*(r;({\bar \alpha}))\cdot\left({\wtilde S}_+\right)^{s+s_0}\rket{0}\ , 
\label{5-7}\\
& &\quad
\left({\wtilde D}'_C{}^*={\wtilde {\cal P}}_{d,\sigma}^*(p;(\lambda),q;(\mu\nu))\left({\wtilde R}_+\right)^{r+r_0}
{\wtilde {\cal Q}}_{d}^*(r;({\bar \alpha}))\right)\nonumber\\
& &\sigma=s+s_0\ . 
\label{5-8}
\eeq
Here, ${\wtilde {\cal P}}_{d,\sigma}^*(p;(\lambda),q;(\mu\nu))$ is defined in the form
\beq\label{5-9}
{\wtilde {\cal P}}_{d,\sigma}^*(p;(\lambda),q;(\mu\nu))=\prod_{k=1}^p{\wtilde {\cal S}}_{\lambda_k,\sigma}^*(d)
\prod_{j=1}^q{\wtilde {\cal S}}_{\mu_j\nu_j,\sigma}^*(d)\ , 
\eeq
\vspace{-0.6cm}
\bsub\label{5-10}
\beq
& &{\wtilde {\cal S}}_{\lambda,\sigma}^*(d)=\frac{1}{2}{\wtilde {\cal S}}_{\alpha\beta,\sigma}^*(d)\ , \qquad (\alpha=\beta=\lambda)
\label{5-10a}\\
& &{\wtilde {\cal S}}_{\mu\nu,\sigma}^*(d)={\wtilde {\cal S}}_{\alpha\beta,\sigma}^*(d)\ . \qquad 
(\alpha=\mu,\ \beta=\nu,\ \mu\neq \nu)
\label{5-10b}
\eeq
\esub
Clearly, the order of the operations is reversed.

Let us consider the connection of the state (\ref{5-7}) to the state (\ref{2-8}). 
We regard the state (\ref{3-1}) as the counterpart of the state (\ref{2-4}): 
\beq\label{5-11}
\rket{\phi_{\sigma}}=\left({\wtilde S}_+\right)^{\sigma}\rket{0}
\longrightarrow 
\rket{\phi}=\exp\left(\frac{v}{u}{\wtilde S}_+\right)\rket{0}\ .
\eeq
The states $\rket{\phi_{\sigma}}$ and $\rket{\phi}$ contain the parameter $\sigma$ and $v$, 
respectively. 
If the correspondence (\ref{5-11}) is accepted, 
$v$ must be relevant to $\sigma$. 
However, 
for the time being, we treat $\sigma$ and $v$ as independent of each other. 
With the use of the relations (\ref{3-9}), (\ref{4-7}) and (\ref{5-2}), we have the 
correspondence 
\beq
& &{\wtilde {\cal D}}_{\alpha,\sigma}^* \longrightarrow 
uw{\maru d}{}_{\alpha}^*\ , \qquad
{\wtilde {\cal D}}_{{\bar \alpha},\sigma}^* \longrightarrow 
uw{\maru d}{}_{{\bar \alpha}}^*\ , 
\label{5-12}\\
& &w=u^2+\frac{\sigma}{2\Omega_0}\ . 
\label{5-13}
\eeq
For example, the case ${\wtilde {\cal D}}_{\alpha,\sigma}^*$ is the following:
\beq\label{5-14}
{\wtilde {\cal D}}_{\alpha,\sigma}^*&\longrightarrow&
\frac{1}{\Omega_0}\left[
-(u{\maru d}{}_{\alpha}^*)\Omega_0(|v|^2-u^2)-\frac{1}{2}(-v{\maru d}{}_\alpha^*)\cdot 2\Omega_0 uv^*\right]
+\frac{\sigma}{2\Omega_0}u{\maru d}{}_{\alpha}^*
\nonumber\\
&=&uw{\maru d}{}_{\alpha}^*\ . 
\eeq
Then, the form (\ref{5-3}) reduces to 
\beq\label{5-15}
{\wtilde {\cal S}}_{\alpha\beta,\sigma}^*(d)\longrightarrow 
u^2w^2({\maru d}{}_{\alpha}^*{\maru d}{}_{\bar \beta}^*-{\maru d}{}_{\bar \alpha}^*{\maru d}{}_{\beta}^*)\ .
\eeq
The relation (\ref{5-15}) gives 
\beq\label{5-16}
{\wtilde {\cal S}}_{\lambda,\sigma}^*(d)\longrightarrow
u^2w^2{\maru d}{}_{\lambda}^*{\maru d}{}_{\bar \lambda}^*\ , \qquad
{\wtilde {\cal S}}_{\mu\nu,\sigma}^*(d)\longrightarrow
u^2w^2({\maru d}{}_{\mu}^*{\maru d}{}_{\bar \nu}^*-{\maru d}{}_{\bar \mu}^*{\maru d}{}_{\nu}^*)\ .
\eeq
Further, we have 
\beq\label{5-17}
& &{\wtilde R}_{\pm,0}={\maru R}_{\pm,0}\ , \nonumber\\
& &{\maru R}_+=\sum_{\alpha}{\maru d}{}_{\alpha}^*{\maru d}{}_{\bar \alpha}\ , \quad
{\maru R}_-=\sum_{\alpha}{\maru d}{}_{\bar \alpha}^*{\maru d}{}_{\alpha}\ , \quad
{\maru R}_0=\frac{1}{2}\sum_{\alpha}({\maru d}{}_{\alpha}^*{\maru d}{}_{\alpha}
-{\maru d}{}_{\bar \alpha}^*{\maru d}{}_{\bar \alpha})\ . 
\eeq
Thus, the state (\ref{5-7}) reduces to 
\beq\label{5-18}
& &\rdket{ss_0:p;(\lambda),q;(\mu\nu),rr_0;({\bar \alpha})}
\rightarrow 
u^n w^{2(p+q)}\rket{p;(\lambda),q;(\mu\nu),rr_0;({\bar \alpha})}\ , 
\nonumber\\
& &\rket{p;(\lambda),q;(\mu\nu),rr_0;({\bar \alpha})}
=\prod_{k=1}^p{\maru d}{}_{\lambda_k}^*{\maru d}{}_{{\bar \lambda}_k}^*
\prod_{j=1}^q({\maru d}{}_{\mu_j}^*{\maru d}{}_{{\bar \nu}_j}^*
-{\maru d}{}_{{\bar \mu}_j}^*{\maru d}{}_{{\nu}_j}^*)
%\nonumber\\
%& &\qquad\qquad\qquad\qquad\qquad
%\times 
\left({\maru R}_+\right)^{r+r_0}\prod_{i=1}^{2r}{\maru d}{}_{{\bar \alpha}_i}^*\rket{\phi}\ .
\nonumber\\
& &
\eeq
The factor $u^nw^{2(p+q)}$ has no meaning, because it disappears by the normalization 
of the state (\ref{5-18}). 
For the normalization, $v$ must be fixed: 
\beq\label{5-19}
\rket{\phi} \sim {\rm the\ normalized}\ \rket{\phi}=u^{2\Omega_0}
\exp\left(\frac{v}{u}{\wtilde S}_+\right)\rket{0}\ . 
\eeq

Our final task of this section is to fix the parameter $v$. 
For this aim, we notice the total fermion and the seniority number operator 
\beq
& &{\maru N}=4\Omega_0|v|^2+
(1-2|v|^2)\sum_{\alpha}({\maru d}{}_{\alpha}^*{\maru d}{}_{\alpha}+{\maru d}{}_{\bar \alpha}^*{\maru d}{}_{\bar \alpha})
\nonumber\\
& &\qquad\qquad\qquad
+2uv\sum_{\alpha}s_{\alpha}{\maru d}{}_{\alpha}^*{\maru d}{}_{\bar \alpha}^*
+2uv^*\sum_{\alpha}s_{\alpha}{\maru d}{}_{\bar \alpha}{\maru d}{}_{\alpha}\ , 
\label{5-20}\\
& &{\maru n}=\sum_{\alpha}({\maru d}{}_{\alpha}^*{\maru d}{}_{\alpha}+{\maru d}{}_{\bar \alpha}^*{\maru d}{}_{\bar \alpha})\ . 
\label{5-21}
\eeq
The total fermion number operator ${\maru N}$ is obtained from 
${\wtilde N}$ by performing the transformation (\ref{2-6b}). 
In the approach B, the seniority number is expressed in terms of the quasiparticle number and, then, 
it may be permitted to set up the form (\ref{5-21}). 
The expectation values of ${\maru N}$ and ${\maru n}$ for the state (\ref{5-18}) are given in the form
\beq\label{5-22}
\langle {\maru N}\rangle=
4\Omega_0|v|^2+(1-2|v|^2)\langle {\maru n}\rangle\ .
\eeq
By requiring the conditions $\langle {\maru N}\rangle=N$ and $\langle {\maru n}\rangle=n$, 
we have 
\beq\label{5-23}
4\Omega_0|v|^2+(1-2|v|^2) n =N\ . 
\eeq 
In the approach B, the condition $\langle {\maru n}\rangle=n=0$ is adopted, i.e., 
the expectation value of ${\maru n}$ for the vacuum of the quasiparticle. 
Then, $|v|^2$ is fixed to 
\beq\label{5-24}
|v|^2=\frac{N}{4\Omega_0}\ , \quad {\rm i.e.,}\quad 
|v|^2=\frac{\sigma}{2\Omega_0}\ .
\eeq
If we follow the idea for the classification of the states for many-fermion system given in (I), 
the set composed from the states (\ref{5-18}) forms a possible orthogonal set and it is 
obtained by making an appropriate linear combination of the states (\ref{2-8}). 
In this way, we can connect the approach A with B through the medium of the 
approach C under the condition $n=0$.

The solution (\ref{5-24}) is a special case of the relation (\ref{5-23}). 
If $\langle {\maru n} \rangle=n\geq 0$ is permitted, the relation 
(\ref{5-23}) gives us 
\beq\label{5-25}
|v|^2=\frac{N-n}{2(2\Omega_0-n)}\ , \quad {\rm i.e.,}\quad
|v|^2=\frac{\sigma}{2s}\ . \qquad \left(s=\Omega_0-\frac{n}{2}\right)
\eeq
Clearly, $|v|^2$ depends on $n$ for a given value of $N$. 
In the approach B, $v$, which is determined under the condition $n=0$, is used 
also for the cases $n>0$. 
This means that the approach B is nothing but a kind of the mean field approximations, 
in which the mean field is applied to any value of $n$. 
The case (\ref{5-25}) may be available for the case where the mean field changes in 
proportion to the change of $n$. 
In this case, $\langle {\wtilde S}_{\pm,0}\rangle$ are given in the form 
\beq\label{5-26}
& &\langle {\wtilde S}_+ \rangle=2\left(\Omega_0-\frac{n}{2}\right)uv^*\ , \qquad
\langle {\wtilde S}_- \rangle=2\left(\Omega_0-\frac{n}{2}\right)uv\ , \nonumber\\
& &\langle {\wtilde S}_0 \rangle=\left(\Omega_0-\frac{n}{2}\right)(|v|^2-u^2)\ . 
\eeq
Of course, for the expectation values (\ref{5-26}), the relation (\ref{5-25}) is 
used. 
The expectation values (\ref{5-26}) can be rewritten as 
\beq\label{5-27}
& &\langle {\wtilde S}_+ \rangle=A^*\sqrt{2s-A^*A}\ , \qquad
\langle {\wtilde S}_- \rangle=A\sqrt{2s-A^*A}\ , \nonumber\\
& &\langle {\wtilde S}_0 \rangle=A^*A-s \ . 
\eeq
Here, $A$ is a new parameter related to $v$:
\beq\label{5-28}
u=\sqrt{1-\frac{A^*A}{2s}}\ , \qquad
v=\frac{A}{\sqrt{2s}}\ , \qquad \sigma=A^*A=\frac{1}{2}(N-n)\ .
\eeq
In next section, we will reconsider the expectation values (\ref{5-27}).

\section{Discussion and concluding remark}

In this paper, we have shown that the approach A is reduced to B under the $c$-number replacement 
for the $su(2)$-generators ${\wtilde S}_{\pm,0}$. 
For this procedure, the role of the ``quasiparticle" is decisive. 
However, we must point out a certain fact which should be examined: 
The idea of the $c$-number replacement should not be applied to the case 
of the operator $\sum_{\lambda}s_{\lambda}{\wtilde {\cal S}}_{\lambda,\sigma}^*(d)$.

We will, first, examine the above fact in an illustrative example $p=q=r=0$, 
i.e., $n=0$. 
The seniority number in this case is $n=0$ and, then, $s=\Omega_0$ and $\sigma=\Omega_0+s_0$. 
Operation of $s_{\lambda}{\wtilde {\cal S}}_{\lambda,\sigma}^*(d)$ on the state 
$\rket{\phi_{\sigma=\Omega_0+s_0}}$ leads to the relation 
\beq\label{6-1}
s_{\lambda}{\wtilde {\cal S}}_{\lambda,\sigma}^*(d)\rket{\phi_{\sigma}}
&=&\left({\wtilde S}_+\right)^{\sigma}s_{\lambda}{\wtilde {\cal S}}_{\lambda}^*(d)\rket{0}
\nonumber\\
&=&\frac{\left(2\Omega_0+\frac{1}{2}\right)(2\Omega_0+1)}{(2\Omega_0)^2}\cdot
\left({\wtilde S}_+\right)^{\sigma}
\left(s_{\lambda}{\tilde c}_{\lambda}^*{\tilde c}_{\bar \lambda}^*-\frac{1}{2\Omega_0}{\wtilde S}_+\right)
\rket{0}\ .
\eeq
Here, we used the relations (\ref{5-4}), (\ref{4-9}), (\ref{2-17a}) and (\ref{2-20}). 
Since $\sum_{\lambda}s_{\lambda}{\tilde c}_{\lambda}^*{\tilde c}_{\bar \lambda}^*={\wtilde S}_+$ and 
$\sum_{\alpha}1=2\Omega_0$, we have
\beq\label{6-2}
\sum_{\lambda}s_{\lambda}{\wtilde {\cal S}}_{\lambda,\sigma}^*(d)\rket{\phi_{\sigma}}
=0\ . \qquad (s=\Omega_0\ , \quad \sigma=\Omega_0+s_0)
\eeq
On the other hand, the correspondences (\ref{5-11}) and (\ref{5-16}) give us 
\beq\label{6-3}
\sum_{\lambda}s_{\lambda}{\wtilde {\cal S}}_{\lambda,\sigma}^*(d)\rket{\phi_{\sigma}}
\longrightarrow 
u^2w^2\sum_{\lambda}s_{\lambda}{\maru d}{}_{\lambda}^*{\maru d}{}_{\bar \lambda}^*\rket{\phi}
\neq 0\ . 
\eeq
The relations (\ref{6-2}) and (\ref{6-3}) do not coincide with each other. 
The above is our examination for the present problem in the case of the 
illustrative example. 
The above conclusion is closely related to the conservation of the fermion number. 
Concerning the fermion number ${\wtilde N}$, we have two relations 
\beq
& &{\wtilde N}\rket{\phi_{\sigma}}=2\sigma\rket{\phi_{\sigma}}\ , 
\label{6-4}\\
& &{\wtilde N}\rket{\phi}=4\Omega_0|v|^2\rket{\phi}+2uv\sum_{\lambda}s_{\lambda}
{\maru d}{}_{\lambda}^*{\maru d}{}_{\bar \lambda}^*\rket{\phi}\ .
\label{6-5} 
\eeq
Because of the relation (\ref{5-24}), we have 
$\rbra{\phi_{\sigma}}{\wtilde N}\rket{\phi_{\sigma}}=\rbra{\phi}{\wtilde N}\rket{\phi}$. 
But, because of the existence of the term $2uv\sum_{\lambda}{\maru d}{}_{\lambda}^*{\maru d}{}_{\bar \lambda}^*$, 
${\wtilde N}\rket{\phi_{\sigma}}$ and ${\wtilde N}\rket{\phi}$ are different from each other.

We will discuss the above problem in general case. 
In (I), we showed the relation 
\beq\label{6-6}
\sum_{\lambda}s_{\lambda}{\wtilde {\cal S}}_{\lambda}^*(c)=-\frac{1}{(2\Omega_0)^2}\left({\wtilde S}_+\right)^2{\wtilde S}_-\ . 
\eeq
Then, $\sum_{\lambda}s_{\lambda}{\wtilde {\cal S}}_{\lambda}^*(d)$ can be expressed as
\beq\label{6-7}
\sum_{\lambda}s_{\lambda}{\wtilde {\cal S}}_{\lambda}^*(d)=\frac{1}{(2\Omega_0)^4}
\left({\wtilde {\mib S}}^2-\Omega_0(\Omega_0+1)\right)\left({\wtilde S}_+\right)^2{\wtilde S}_-\ . 
\eeq
Since ${\wtilde S}_-$ is present in the right-sides of the relations (\ref{6-6}) and (\ref{6-7}), 
both operators can be essentially regarded as vanishing operators 
in the minimum weight state. 
From this mention, we have 
\bsub\label{6-8}
\beq
& &\sum_{\lambda}s_{\lambda}{\wtilde {\cal S}}_{\lambda}^*(d)
\rdket{p;(\lambda),q;(\mu\nu),rr_0;({\bar \alpha})}=0\ , 
\label{6-8a}\\
{\rm i.e.,}\quad 
& &\left({\wtilde S}_+\right)^{s+s_0}\sum_{\lambda}s_{\lambda}{\wtilde {\cal S}}_{\lambda}^*(d)
\rdket{p;(\lambda),q;(\mu\nu),rr_0;({\bar \alpha})}=0\ . 
\label{6-8b}
\eeq
\esub
Here, we used the relation (\ref{4-14}). 
The relation (\ref{6-8b}) gives us 
\beq\label{6-9}
\sum_{\lambda}s_{\lambda}{\wtilde {\cal S}}_{\lambda,\sigma}^*(d)
\rdket{ss_0:p;(\lambda),q;(\mu\nu),rr_0;({\bar \alpha})}=0\ . \qquad (\sigma=s+s_0)
\eeq
In the approach B, generally, we have 
\beq\label{6-10}
\sum_{\lambda}s_{\lambda}{\maru d}{}_{\lambda}^*{\maru d}{}_{\bar \lambda}^*
\rket{p;(\lambda),q;(\mu\nu),rr_0;({\bar \alpha})}\neq 0\ .
\eeq
Here, we used the relation (\ref{5-16}). 
In the present general case, the relation (\ref{6-9}) and (\ref{6-10}) also do not coincide.

In more detail, we will investigate the properties of the operator 
$\sum_{\lambda}s_{\lambda}{\wtilde {\cal S}}_{\lambda,\sigma}^*(d)$. 
After rather lengthy calculation, the explicit form can be derived: 
\beq\label{6-11}
\sum_{\lambda}s_{\lambda}{\wtilde {\cal S}}_{\lambda,\sigma}^*(d)=
\frac{1}{(2\Omega_0)^4}{\wtilde S}_+
\left({\wtilde {\mib S}}^2-\Omega_0(\Omega_0+1)\!\right)\left(
{\wtilde {\mib S}}^2-(\!\sigma-{\wtilde S}_0)
\left(\!\sigma-{\wtilde S}_0+1\!\right)\!\right) .\qquad
\eeq
It satisfies the relation 
\beq\label{6-12}
[\ {\wtilde {\mib S}}^2\ , \ \sum_{\lambda}s_{\lambda}{\wtilde {\cal S}}_{\lambda,\sigma}^*(d)\ ]=0\ , \quad
[\ {\wtilde S}_0\ , \ \sum_{\lambda}s_{\lambda}{\wtilde {\cal S}}_{\lambda,\sigma}^*(d)\ ]=
\sum_{\lambda}s_{\lambda}{\wtilde {\cal S}}_{\lambda,\sigma}^*(d)\ .
\eeq
The relation (\ref{6-12}) tells us that $\sum_{\lambda}s_{\lambda}{\wtilde {\cal S}}_{\lambda,\sigma}^*(d)$ can not 
play the role of building block for composing the orthogonal basis. 
Certainly, we have 
\beq\label{6-13}
& &\sum_{\lambda}s_{\lambda}{\wtilde {\cal S}}_{\lambda,\sigma}^*(d)
\rdket{ss_0:p;(\lambda),q;(\mu\nu),rr_0;({\bar \alpha})}
\nonumber\\
&=&(s-\Omega_0)(s+\Omega_0+1)(s+s_0-\sigma)(s-s_0+\sigma+1)/(2\Omega_0)^4 \nonumber\\
& &\times
\rdket{ss_0+1:p;(\lambda),q;(\mu\nu),rr_0;({\bar \alpha})}\ .
\eeq
In the case $s=\Omega_0$ or $\sigma=s+s_0$ ($s=-(\Omega_0+1)<0$, $\sigma=-(s-s_0+1)<0$), 
the right-hand side of the relation (\ref{6-13}) vanishes and in other cases, 
the role of $\sum_{\lambda}s_{\lambda}{\wtilde {\cal S}}_{\lambda,\sigma}^*(d)$ is 
equivalent to that of ${\wtilde S}_+$: 
$s_0$ changes to $s_0+1$. 
Naturally, the seniority number does not change. 
Next, we contact with the conservation of the fermion number. 
In the present case, the relations similar to the relations (\ref{6-4}) and (\ref{6-5}) are derived: 
\beq
& &{\wtilde N}\rdket{ss_0:p;(\lambda),q;(\mu\nu),rr_0;({\bar \alpha})}
=(2\sigma+n)\rdket{ss_0:p;(\lambda),q;(\mu\nu),rr_0;({\bar \alpha})}\ , 
\label{6-14}\\
& &{\maru N}\rket{p;(\lambda),q;(\mu\nu),rr_0;({\bar \alpha})}
=(2(2\Omega_0-n)|v|^2+n)\rket{p;(\lambda),q;(\mu\nu),rr_0;({\bar \alpha})}\nonumber\\
& &\qquad\qquad\quad
+(2uv\sum_{\lambda}s_{\lambda}{\maru d}{}_{\lambda}^*{\maru d}{}_{\bar \lambda}^*
+2uv^*\sum_{\lambda}s_{\lambda}{\maru d}{}_{\bar \lambda}{\maru d}{}_{\lambda})
\rket{p;(\lambda),q;(\mu\nu),rr_0;({\bar \alpha})}\ .\quad
\label{6-15}
\eeq
If we follow the condition (\ref{5-25}), we have $2\sigma+n=2(2\Omega_0-n)|v|^2+n=N$. 
However, in the relation (\ref{6-15}), there exist the terms related to 
$\sum_{\lambda}s_{\lambda}{\maru d}{}_{\lambda}^*{\maru d}{}_{\bar \lambda}^*$ and 
$\sum_{\lambda}s_{\lambda}{\maru d}{}_{\bar \lambda}{\maru d}_{\lambda}$. 
This feature comes from the non-conservation of the fermion number and it is 
in the same situation as that in the illustrative example.

The expressions (\ref{6-14}) and (\ref{6-15}) coincide, if the 
following condition is permitted: 
\beq\label{6-16}
\sum_{\lambda}s_{\lambda}{\maru d}{}_{\lambda}^*{\maru d}{}_{\bar \lambda}^*=0\ , \qquad
\sum_{\lambda}s_{\lambda}{\maru d}{}_{\bar \lambda}{\maru d}{}_{\lambda}=0\ .
\eeq
However, it may be self-evident that the condition (\ref{6-16}) cannot be accepted. 
With the aid of the Dirac's canonical theory for the constraint system, the 
problem can be expected to be solved. 
As a set of operators which corresponds to the set of the fermions 
${\maru d}{}_{\alpha}^*$ etc., we introduce the following new set: 
\beq\label{6-17}
{\maru d}{}_{\alpha}^*\longrightarrow {\hat d}_{\alpha}^*\ , \quad
{\maru d}{}_{\bar \alpha}^*\longrightarrow {\hat d}_{\bar \alpha}^*\ , \quad
{\maru d}{}_{\alpha}\longrightarrow {\hat d}_{\alpha}\ , \quad
{\maru d}{}_{\bar \alpha}\longrightarrow {\hat d}_{\bar \alpha}\ .
\eeq
Further, we require the condition
\beq\label{6-18}
{\hat \chi}_+=\sum_{\alpha}s_{\alpha}{\hat d}_{\alpha}^*{\hat d}_{\bar \alpha}^* \approx 0\ , \qquad
{\hat \chi}_-=\sum_{\alpha}s_{\alpha}{\hat d}_{\bar \alpha}{\hat d}_{\alpha} \approx 0\ .
\eeq
Here, the symbol $\approx$ denotes the equal sign in the sense of the Dirac theory. 
If we require the condition (\ref{6-18}) as the constraint, the Dirac theory gives us 
the following anti-commutation relation: 
\beq\label{6-19}
& &\{\ {\hat d}_{\alpha}\ , \ {\hat d}_{\beta}^*\ \}=\delta_{\alpha\beta}
-s_{\alpha}{\hat d}_{\bar \alpha}^*\left(2{\hat S}\right)^{-1}s_{\beta}{\hat d}_{\bar \beta}\ , \quad
\{\ {\hat d}_{\alpha}\ , \ {\hat d}_{\beta}\ \}=0\ , 
\nonumber\\
& &\{\ {\hat d}_{\bar \alpha}\ , \ {\hat d}_{\bar \beta}^*\ \}
=\delta_{\alpha\beta}-s_{\alpha}{\hat d}_{\alpha}^*\left(2{\hat S}\right)^{-1}s_{\beta}{\hat d}_{\beta}\ , 
\qquad
\{\ {\hat d}_{\bar \alpha}\ , \ {\hat d}_{\bar \beta}\ \}=0\ , 
\nonumber\\
& &\{\ {\hat d}_{\alpha}\ , \ {\hat d}_{\bar \beta}^*\ \}=
s_{\alpha}{\hat d}_{\bar \alpha}^*\left(2{\hat S}\right)^{-1}s_{\beta}{\hat d}_{\beta}\ , \quad
\{\ {\hat d}_{\alpha}\ , \ {\hat d}_{\bar \beta}\ \}=0\ . 
\eeq
Here ${\hat S}$ denotes 
\beq\label{6-20}
{\hat S}=\Omega_0-\frac{1}{2}{\hat n}\ , \qquad
{\hat n}=\sum_{\alpha}({\hat d}_{\alpha}^*{\hat d}_{\alpha}
+{\hat d}_{\bar \alpha}^*{\hat d}_{\bar \alpha})\ . 
\eeq
Under the condition (\ref{6-18}), the number of the degrees of freedom 
decreases by 2. 
In order to recover this decrease, we regard the parameters 
$(A, A^*)$ defined in the relation (\ref{5-28}) as boson 
operators $({\hat A}, {\hat A}^*)$ which 
commute with any of $({\hat d}_{\alpha}, {\hat d}_{\alpha}^*, {\hat d}_{\bar \alpha}, {\hat d}_{\bar \alpha}^*)$. 
In this case, the $c$-numbers $S_{\pm,0}$ given in the 
relation (\ref{5-27}) become the $q$-numbers ${\hat S}_{\pm,0}$: 
\bsub\label{6-21}
\beq
& &{\hat S}_+={\hat A}^*\cdot\sqrt{2{\hat S}-{\hat A}^*{\hat S}}\ , \quad
{\hat S}_-=\sqrt{2{\hat S}-{\hat A}^*{\hat A}}\cdot {\hat A}\ , \quad
{\hat S}_0={\hat A}^*{\hat A}-{\hat S}\ , 
\label{6-21a}\\
& &[\ {\hat S}_+\ , \ {\hat S}_-\ ]=2{\hat S}_0\ , \qquad
[\ {\hat S}_0\ , \ {\hat S}_{\pm}\ ]=\pm{\hat S}_{\pm}\ . 
\label{6-21b}
\eeq
\esub
The above is nothing but the Holstein-Primakoff boson representation 
for the $su(2)$-algebra. 
We can see that ${\hat S}$ and ${\hat n}$ defined in the relation (\ref{6-20}) denote 
the magnitude of the $S$-spin and the seniority number, respectively. 

The above argument permits us to intend to formulate the present model in the 
space constructed by the boson operators $({\hat A},{\hat A}^*)$ and the set of 
the operators $({\hat d}_{\alpha},{\hat d}_{\alpha}^*)$ obeying the 
anti-commutation relation (\ref{6-19}). 
We call this space as the boson-fermion space. 
Formally, the boson-fermion space is larger than the original 
fermion space. 
But, the condition that $(2{\hat S}-{\hat A}^*{\hat A})$ should be positive-definite 
and the constraint (\ref{6-18}) guarantee the one-to-one correspondence between 
the both spaces. 
For the convenience of later discussion, 
we give a comment on the relations related to the constraint 
(\ref{6-18}). 
For example, we have 
\bsub\label{6-22}
\beq
& &[\ {\hat \chi}_-\ , \ {\hat d}_{\alpha}^*\ ]=2{\hat \chi}_+\left(2{\hat S}\right)^{-1}{\hat \chi}_-\cdot
s_{\alpha}{\hat d}_{\bar \alpha} \approx 0\ , 
\label{6-22a}\\
& &[\ {\hat n}\ , \ {\hat d}_{\alpha}^*\ ]={\hat d}_{\alpha}^*-2{\hat \chi}_+\left(2{\hat S}\right)^{-1}\cdot
s_{\alpha}{\hat d}_{\bar \alpha} \approx {\hat d}_{\alpha}^*\ . 
\label{6-22b}
\eeq
\esub
Here, we used the relation (\ref{6-18}). 
We remove the terms directly related to ${\hat \chi}_{\pm}$ from the above two relations and 
use the equal 
sign $\approx$. 
Hereafter, any relation will be expressed in the form, in which the terms directly related to 
${\hat \chi}_{\pm}$ are removed. 
Let the counterpart of $({\tilde c}_{\alpha}^*, {\tilde c}_{\bar \alpha}^*)$ be expressed as 
\beq\label{6-23}
{\tilde c}_{\alpha}^* \longrightarrow {\bar c}_{\alpha}^*\ , \qquad
{\tilde c}_{\bar \alpha}^* \longrightarrow {\bar c}_{\bar \alpha}^*\ .
\eeq
Since ${\tilde c}_{\alpha}^*$ and ${\tilde c}_{\bar \alpha}^*$ are expressed in the form 
(\ref{2-6b}), we define ${\bar c}_{\alpha}^*$ and ${\bar c}_{\bar \alpha}^*$ 
in the following form: 
\beq\label{6-24}
& &{\tilde c}_{\alpha}^*=u{\maru d}{}_{\alpha}^*+v^*s_{\alpha}{\maru d}{}_{\bar \alpha}
\longrightarrow 
{\bar c}_{\alpha}^*={\hat d}_{\alpha}^*{\hat u}+{\hat v}^*s_{\alpha}{\hat d}_{\bar \alpha}\ , 
\nonumber\\
& &{\tilde c}_{\bar \alpha}^*=u{\maru d}{}_{\bar \alpha}^*-v^*s_{\alpha}{\maru d}{}_{\alpha}
\longrightarrow 
{\bar c}_{\bar \alpha}^*={\hat d}_{\bar \alpha}^*{\hat u}-{\hat v}^*s_{\alpha}{\hat d}_{\alpha}\ .
\eeq
From the relation (\ref{5-28}), ${\hat u}$ and ${\hat v}$ are defined in the form 
\beq\label{6-25}
u\longrightarrow {\hat u}=\sqrt{1-\frac{{\hat A}^*{\hat A}}{2{\hat S}}}\ , \qquad
v\longrightarrow {\hat v}=\frac{\hat A}{\sqrt{2{\hat S}}}\ . 
\eeq
Then, we can derive the anti-commutation relation 
\beq\label{6-26}
\{\ {\bar c}_{\alpha}\ , \ {\bar c}_{\bar \beta}^*\ \}\approx \delta_{\alpha\beta}\ , \qquad
\{\ {\bar c}_{\alpha}\ , \ {\bar c}_{\beta}\ \}=0 \ . 
\eeq
The counterpart of ${\wtilde S}_{\pm,0}$ defined in the relation (\ref{2-1}) is given 
\beq\label{6-27}
& &{\wtilde S}_+\longrightarrow {\ovl S}_+=\sum_{\alpha}s_{\alpha}{\bar c}_{\alpha}^*{\bar c}_{\bar \alpha}^*\approx {\hat S}_+\ , 
\nonumber\\
& &{\wtilde S}_-\longrightarrow {\ovl S}_-=\sum_{\alpha}s_{\alpha}{\bar c}_{\bar \alpha}{\bar c}_{\alpha}\approx {\hat S}_-\ , 
\nonumber\\
& &{\wtilde S}_0\longrightarrow {\ovl S}_0
=\frac{1}{2}\sum_{\alpha}({\bar c}_{\alpha}^*{\bar c}_{\alpha}+{\bar c}_{\bar \alpha}^*{\bar c}_{\bar \alpha})
\approx {\hat S}_0\ . 
\eeq
Of course, ${\hat S}_{\pm,0}$ in the relation (\ref{6-27}) is given in the form (\ref{6-21}). 
The set $({\ovl S}_{\pm,0})$ satisfies 
\beq\label{6-28}
[\ {\ovl S}_+\ , \ {\ovl S}_-\ ]\approx 2{\ovl S}_0\ , \qquad
[\ {\ovl S}_0\ , \ {\ovl S}_{\pm}\ ]\approx \pm {\ovl S}_{\pm}\ .
\eeq
The counterpart of ${\wtilde R}_{\pm,0}$ defined in the relation (\ref{2-21}) is given as 
\beq\label{6-29}
& &{\wtilde R}_+\longrightarrow {\ovl R}_+=\sum_{\alpha}{\bar c}_{\alpha}^*{\bar c}_{\bar \alpha}\approx {\hat R}_+\ , 
\nonumber\\
& &{\wtilde R}_-\longrightarrow {\ovl R}_-=\sum_{\alpha}{\bar c}_{\bar \alpha}^*{\bar c}_{\alpha}\approx {\hat R}_-\ , 
\nonumber\\
& &{\wtilde R}_0\longrightarrow {\ovl R}_0
=\frac{1}{2}\sum_{\alpha}({\bar c}_{\alpha}^*{\bar c}_{\alpha}-{\bar c}_{\bar \alpha}^*{\bar c}_{\bar \alpha})
\approx{\hat R}_0\ . 
\eeq
Here, ${\hat R}_{\pm,0}$ are defined as 
\beq\label{6-30}
{\hat R}_+=\sum_{\alpha}{\hat d}_{\alpha}^*{\hat d}_{\bar \alpha}\ , \qquad
{\hat R}_-=\sum_{\alpha}{\hat d}_{\bar \alpha}^*{\hat d}_{\alpha}\ , \qquad
{\hat R}_0=\frac{1}{2}\sum_{\alpha}({\hat d}_{\alpha}^*{\hat d}_{\alpha}-{\hat d}_{\bar \alpha}^*{\hat d}_{\bar \alpha})\ .
\eeq
We can prove the relations 
\beq
& &[\ {\hat R}_+\ , \ {\hat R}_-\ ]=2{\hat R}_0\ , \qquad
[\ {\hat R}_0\ , \ {\hat R}_{\pm}\ ]=\pm {\hat R}_{\pm}\ , 
\label{6-32}\\
& &[ \ {\hat n}\ , \ {\hat R}_{\pm,0}\ ] =0\ , \quad {\rm i.e.,}\quad
[\ {\hat S}\ , \ {\hat R}_{\pm,0}\ ]=0 \ .
\label{6-32}
\eeq
The relation (\ref{6-32}) gives us 
\beq\label{6-33}
[\ {\rm any\ of}\ {\hat R}_{\pm,0}\ , \ {\rm any\ of}\ {\hat S}_{\pm,0}\ ]=0\ . 
\eeq
The operators ${\ovl R}_{\pm,0}$ satisfy
\beq\label{6-34}
[\ {\ovl R}_+\ , \ {\ovl R}_-\ ]\approx 2{\ovl R}_0\ , \qquad
[\ {\ovl R}_0\ , \ {\ovl R}_{\pm}\ ]\approx \pm{\ovl R}_{\pm}\ .
\eeq

It may be interesting to investigate the ``quasiparticle" introduced in \S 3 
in the framework of the above formalism. 
The counterpart of $({\tilde d}_{\alpha +}^*, {\tilde d}_{\alpha -}^*, {\tilde d}_{{\bar \alpha}+}^*, 
{\tilde d}_{{\bar \alpha}-}^*)$ defined in the relation (\ref{3-3}) is given in the 
following form: 
\beq\label{6-35}
& &{\tilde d}_{\alpha +}^*
\longrightarrow 
{\bar d}_{\alpha +}^*
=\frac{1}{2\Omega_0}\left[\left(\Omega_0-{\ovl S}_0\right){\bar c}_{\alpha}^*
-s_{\alpha}{\bar c}_{\bar \alpha}{\ovl S}_+\right]\approx {\hat d}_{\alpha +}^*\ , \nonumber\\
& &{\tilde d}_{\alpha -}^*
\longrightarrow 
{\bar d}_{\alpha -}^*
=\frac{1}{2\Omega_0}\left[\left(\Omega_0+{\ovl S}_0\right)s_{\alpha}{\bar c}_{\bar \alpha}
-{\bar c}_{\alpha}^*{\ovl S}_-\right]\approx {\hat d}_{\alpha -}^*\ , \nonumber\\
& &{\tilde d}_{{\bar \alpha} +}^*
\longrightarrow 
{\bar d}_{{\bar \alpha} +}^*
=\frac{1}{2\Omega_0}\left[\left(\Omega_0-{\ovl S}_0\right){\bar c}_{\bar \alpha}^*
+s_{\alpha}{\bar c}_{\alpha}{\ovl S}_+\right]\approx {\hat d}_{{\bar \alpha} +}^*\ , \nonumber\\
& &{\tilde d}_{{\bar \alpha} -}^*
\longrightarrow 
{\bar d}_{{\bar \alpha} -}^*
=\frac{1}{2\Omega_0}\left[-\left(\Omega_0+{\ovl S}_0\right)s_{\alpha}{\bar c}_{\alpha}
-{\bar c}_{\bar \alpha}^*{\ovl S}_-\right]\approx {\hat d}_{{\bar \alpha} -}^*\ .
\eeq
Here ${\hat d}_{\alpha +}^*$ etc. are expressed in the form 
\beq\label{6-36}
& &
{\hat d}_{\alpha +}^*
=\left(\frac{\Omega_0+{\hat S}+1}{2\Omega_0}\right){\hat d}_{\alpha}^*{\hat u}
+\left(\frac{\Omega_0-{\hat S}}{2\Omega_0}\right){\hat v}^*s_{\alpha}{\hat d}_{\bar \alpha}\ , 
\nonumber\\
& &
{\hat d}_{\alpha -}^*
=-\left(\frac{\Omega_0+{\hat S}+1}{2\Omega_0}\right){\hat d}_{\alpha}^*{\hat v}
+\left(\frac{\Omega_0-{\hat S}}{2\Omega_0}\right){\hat u}s_{\alpha}{\hat d}_{\bar \alpha}\ , 
\nonumber\\
& &
{\hat d}_{{\bar \alpha} +}^*
=\left(\frac{\Omega_0+{\hat S}+1}{2\Omega_0}\right){\hat d}_{\bar \alpha}^*{\hat u}
-\left(\frac{\Omega_0-{\hat S}}{2\Omega_0}\right){\hat v}^*s_{\alpha}{\hat d}_{\alpha}\ , 
\nonumber\\
& &
{\hat d}_{{\bar \alpha} -}^*
=-\left(\frac{\Omega_0+{\hat S}+1}{2\Omega_0}\right){\hat d}_{\bar \alpha}^*{\hat v}
-\left(\frac{\Omega_0-{\hat S}}{2\Omega_0}\right){\hat u}s_{\alpha}{\hat d}_{\alpha}\ . 
\eeq
The counterpart of $\rket{\phi_{\sigma}}$ is given as 
\beq\label{6-37}
\rket{\phi_{\sigma}}=\left({\hat S}_+\right)^{\sigma}\rket{0}
\longrightarrow 
\ket{\phi_{\sigma}}=\sqrt{\frac{(2\Omega_0)!}{(2\Omega_0-\sigma)!}}\left({\hat A}^*\right)^{\sigma}\ket{0}\ . 
\eeq
The relation which corresponds to the relation (\ref{3-2}) is derived: 
\beq\label{6-38}
{\hat d}_{\alpha +}\ket{\phi_{\sigma}}={\hat d}_{\alpha -}\ket{\phi_{\sigma}}
={\hat d}_{{\bar \alpha} +}\ket{\phi_{\sigma}}={\hat d}_{{\bar \alpha} -}\ket{\phi_{\sigma}}
=0\ .
\eeq
Here, we used the relation 
\beq\label{6-39}
{\hat d}_{\alpha}\ket{\phi_{\sigma}}={\hat d}_{\bar \alpha}\ket{\phi_{\sigma}}=0\ , \quad
{\rm i.e.,}\quad
{\hat S}\ket{\phi_{\sigma}}=\Omega_0\ket{\phi_{\sigma}}\ . 
\eeq
We presented a possible formalism for treating the pairing correlation in the 
framework of the conservation of the fermion number. 
In this sense, it is connected with the approach C. 
It may be interesting to see that this form is analogous to the approach B. 
In a sense, it can be regarded as a revised version of the approach B.

In this paper, we discussed the relation between two approaches 
A and B. 
Both are apparently quite different from each other. 
But, our idea called the approach C clarified that they are connected other 
under the $c$-number replacement of ${\wtilde S}_{\pm,0}$. 
If we do not adopt the $c$-number replacement, the approach C and the form presented 
in this section may be equivalent. 
For the classification, the introduction of the ``quasiparticle" is decisive. 
However, at the present stage, we must mention that the use of the 
``quasiparticle" is very tedious for the practical calculation. 
Therefore, it may be desirable to modify the present form of the ``quasiparticle", if it is 
useful for the practical purpose.

\vspace{1cm}
\section*{Acknowledgement}

One of the authors (Y.T.) 
is partially supported by the Grants-in-Aid of the Scientific Research 
(No.23540311) from the Ministry of Education, Culture, Sports, Science and 
Technology in Japan.

%\appendix
%\section{General formula for deriving the relation (\ref{4-16})}

\end{document}